\documentclass[conference]{IEEEtran}
\IEEEoverridecommandlockouts
\usepackage{cite}
\usepackage{amsmath,amssymb,amsfonts}
\usepackage{algorithmic}
\usepackage{graphicx}
\usepackage{textcomp}
\usepackage{xcolor}

\usepackage{multirow}
\usepackage{hhline}
\usepackage{subcaption}
\usepackage{hyperref}
\usepackage{soul}
\usepackage[numbers]{natbib}

\def\BibTeX{{\rm B\kern-.05em{\sc i\kern-.025em b}\kern-.08em
    T\kern-.1667em\lower.7ex\hbox{E}\kern-.125emX}}
\begin{document}

\title{The ISC Creator: Human-Centered Design of Learning Analytics Interactive Indicator Specification Cards}

\author{
    %\IEEEauthorblockN{1\textsuperscript{st} Shoeb Joarder}
    \IEEEauthorblockN{Shoeb Joarder}
    \IEEEauthorblockA{
        \textit{Social Computing Group} \\
        \textit{Faculty of Computer Science} \\
        \textit{University of Duisburg-Essen}\\
        Duisburg, Germany \\
        shoeb.joarder@uni-due.de
    }
    ~\\
    \and
    %\IEEEauthorblockN{2\textsuperscript{nd} Mohamed Amine Chatti}
    \IEEEauthorblockN{Mohamed Amine Chatti}
    \IEEEauthorblockA{
        \textit{Social Computing Group} \\
        \textit{Faculty of Computer Science} \\
        \textit{University of Duisburg-Essen}\\
        Duisburg, Germany \\
        mohamed.chatti@uni-due.de
    }
}

\maketitle

\begin{abstract}
  Emerging research on human-centered learning analytics (HCLA) has demonstrated the importance of involving diverse stakeholders in co-designing learning analytics (LA) systems. However, there is still a demand for effective and efficient methods to co-design LA dashboards and indicators. Indicator Specification Cards (ISCs) have been introduced recently to facilitate the systematic co-design of indicators by different LA stakeholders. In this paper, we strive to enhance the user experience and usefulness of the ISC-based indicator design process. Towards this end, we present the systematic design, implementation, and evaluation details of the ISC Creator, an interactive LA tool that allows low-cost and flexible design of LA indicators. Our findings demonstrate the importance of carefully considered interactivity and recommendations for orienting and supporting non-expert LA stakeholders to design custom LA indicators.
\end{abstract}


%%%%%%%%%%%%%%%%%%%%%%%%%%%%%%%%%%%%%%%
\section{Introduction}
%%%%%%%%%%%%%%%%%%%%%%%%%%%%%%%%%%%%%%%
Learning analytics (LA) systems typically present learners' interaction data gathered from various learning environments through indicator visualizations on dashboards. Although numerous LA dashboards and indicators have been suggested, we still lack evidence that these have any evident impact on learning and teaching. This hinders the acceptance and adoption of LA dashboards and indicators at scale in schools, universities, and workplaces. Along with technical issues, there are more crucial pedagogical and methodological problem areas related to designing LA dashboards and indicators. These include a lack of adherence to human-computer interaction (HCI) and information visualization (InfoVis) guidelines and theories \cite{gavsevic2015let, jivet2018license, klerkx2017learning}. But, the most important reason is that most LA dashboards and indicators are not adopted by the end users because they are not well aligned with users' needs and expectations \cite{chatti2019perla, gavsevic2016we}. Human-centered learning analytics (HCLA) has recently emerged as an approach emphasizing human factors in LA and promoting HCI in LA. This approach aims at having LA stakeholders in the loop and involving them throughout the LA process, which has been shown as key to increasing user acceptance and adoption of LA systems \cite{buckingham2019human, chatti2021designing, dimitriadis2021human, sarmiento2022participatory}. Recently, researchers have experimented with using LA cards to co-design LA tools \cite{alvarez2020deck} and LA indicators \cite{chatti2020design}. However, in their study, the authors in \cite{chatti2021designing} experienced that using Indicator Specification Cards (ISCs) (see Figure \ref{fig:isc-example} for an example) to co-design LA indicators was a complex, time-consuming, and resource-intensive task.
% Current research on HCLA highlights the benefits of active user involvement in the LA design process and shows successful co-design of LA tools with different stakeholders \cite{dollinger2019working, holstein2019co}. Prior work has also demonstrated that non-expert LA stakeholders benefit from LA cards to co-design LA tools \cite{alvarez2020deck}. However, most of this research primarily focused on the participatory design of LA tools and platforms (macro design level) rather than the systematic design of underlying indicators (micro design level). To address this research gap, \citet{chatti2020design} proposed the Human-Centered Indicator Design (HCID) approach and Indicator Specification Cards (ISC) to facilitate the systematic co-design of LA indicators. In another study, the authors experienced that using ISCs was a complex, time-consuming, and resource-intensive task \cite{chatti2021designing}. 
Building upon the idea of providing ISCs to help non-expert LA stakeholders co-design LA indicators, as proposed in \cite{chatti2020design}, in this work, we highlight and address the limitations in the current design of ISCs. Our investigation consists of initial interviews with non-expert LA stakeholders to identify the limitations of current ISCs. To address these limitations, we follow the human-centered design (HCD) approach \cite{norman2013design} to systematically design and implement the \textit{ISC Creator}, an interactive LA tool that allows intuitive, low-cost, and flexible design of LA indicators. Using a semi-structured interview format and a think-aloud protocol, we evaluated the \textit{ISC Creator} regarding user acceptance and satisfaction based on the technology acceptance model (TAM) \cite{davis1989perceived}. The results of our study provide qualitative evidence that it is essential to scaffold LA stakeholders with interactivity mechanisms and theoretically sound recommendations to design custom LA indicators. 

The remainder of this paper is structured as follows. First, we outline the theoretical background of this research and discuss related work in Section \ref{background}. In Section \ref{hcd-isc}, we discuss the human-centered design and implementation of the \textit{ISC Creator}. Next, we describe the user study, present the results, and discuss our findings in Section \ref{evaluation}). Afterward, we discuss the limitations of the work in Section \ref{limitations}. Finally, in Section \ref{conclusion}, we conclude the paper with a summary and avenues for future work. 
%--- However, the application of xx is still under-investigated in current LA research and practice.
% The primary focus of this work is to explore how the ISC Creator can achieve a practical HCLA
% approach and improve the acceptance of and satisfaction with LA systems. Toward this end, we present the design and implementation details of the ISC Creator. This intuitive tool allows different LA stakeholders to design custom LA indicators that meet their needs.
%%%%%%%%%%%%%%%%%%%%%%%%%%%%%%%%%%%%%%%
\section{Background and Related Work}
\label{background}
Human-centered learning analytics (HCLA) prioritizes human factors in learning analytics (LA). It aims at incorporating human-computer interaction (HCI) to involve stakeholders throughout the entire LA process \cite{chatti2019perla, chatti2021designing}. Engaging humans in designing, developing, and evaluating LA systems aims to effectively serve the needs of diverse LA stakeholders and their various goals \cite{buckingham2019human}. To achieve HCLA, HCI approaches like design thinking, human-centered design (HCD), participatory design, co-design, and value-sensitive design can be applied to LA \cite{sarmiento2022participatory}. In this line of research, user involvement in the LA design process is encouraged, and successful co-design processes for LA tools with different stakeholders have been demonstrated, e.g., \cite{dollinger2019working, holstein2019co, alvarez2020deck}. Prior work has also shown that non-expert LA stakeholders benefit from LA cards to co-design LA tools \cite{alvarez2020deck}. However, existing case studies have mainly focused on the participatory design of LA tools and platforms (macro design level) instead of the systematic design of the underlying indicators (micro design level). To address this gap, \citet{chatti2020design} introduced human-centered indicator design (HCID) as an HCLA approach that involves users in the systematic design of LA indicators that meet their needs using Indicator Specification Cards (ISCs). 

\begin{figure}[!ht]
	\centering
    \includegraphics[width=0.49\textwidth]{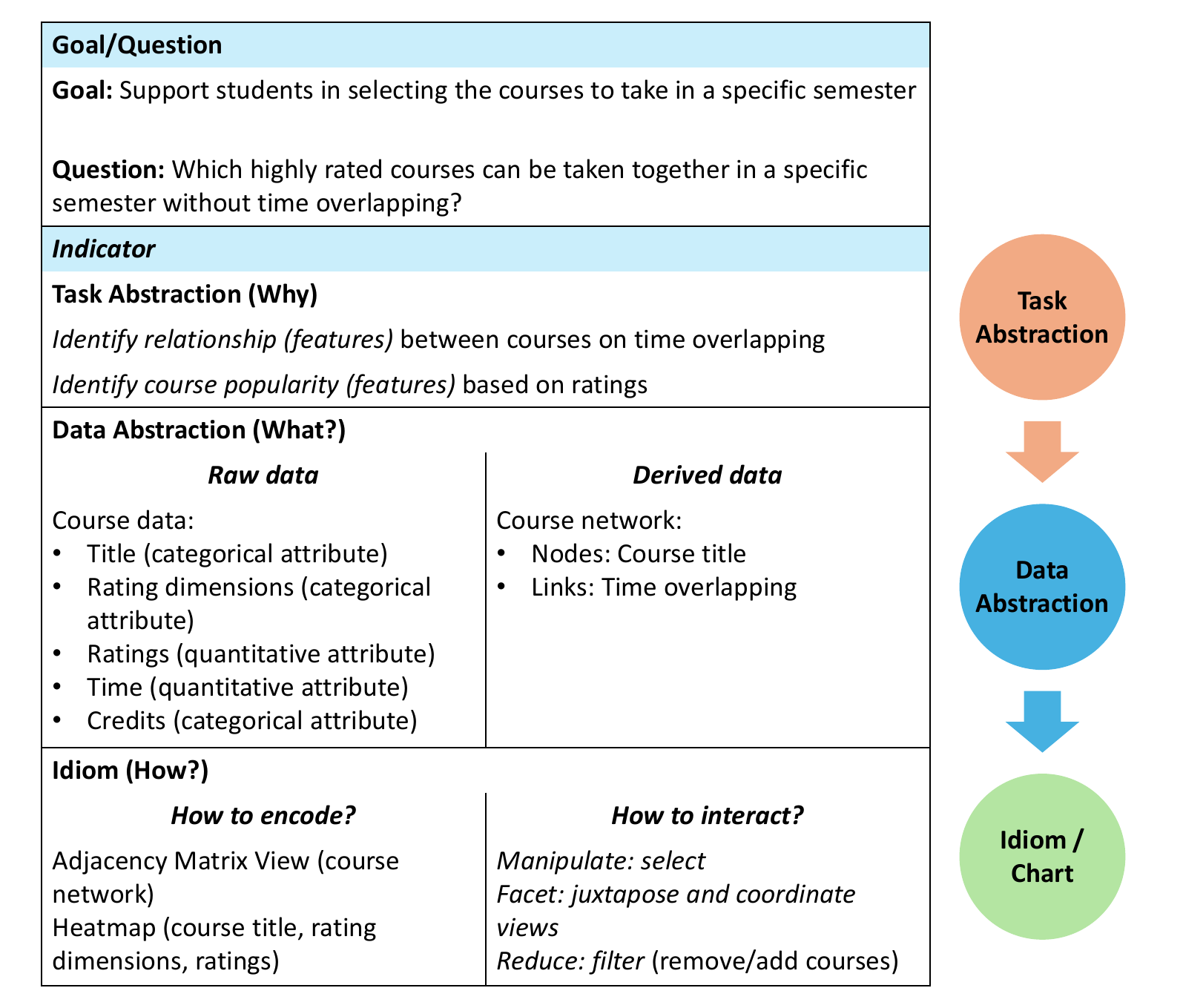}
    \caption{Indicator Specification Card (ISC) example \cite{chatti2020design}}
    \label{fig:isc-example}
\end{figure}

\begin{figure*}
    \begin{center}
        \begin{subfigure}[normla]{0.49\textwidth}
            \centering
            \includegraphics[width=1\textwidth]{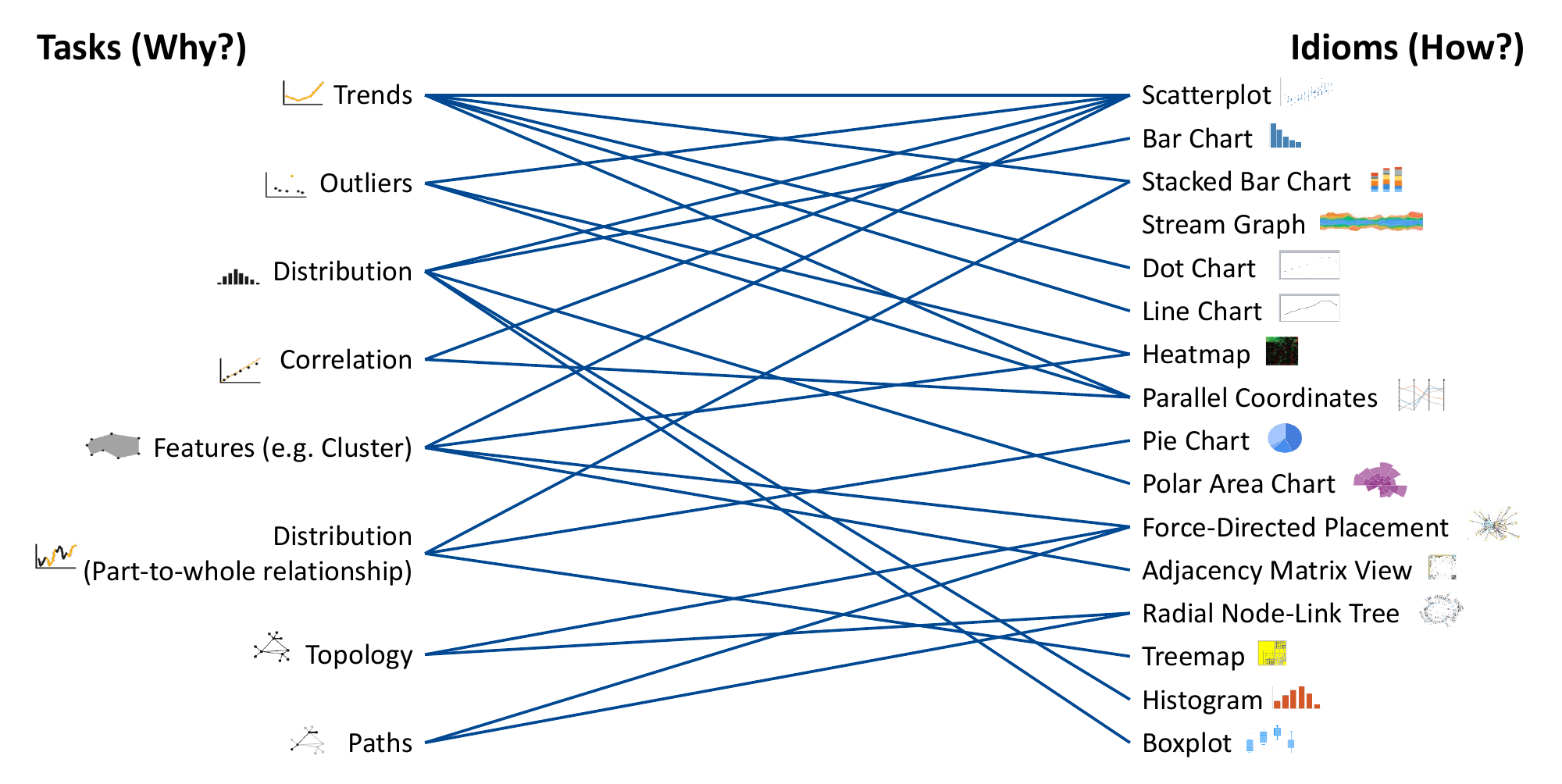}
            \caption{Why? $\to$ How? }
            \label{subfig:infovis-mapping-a}
        \end{subfigure}
        ~
        \begin{subfigure}[normla]{0.49\textwidth}
            \centering
            \includegraphics[width=1\textwidth]{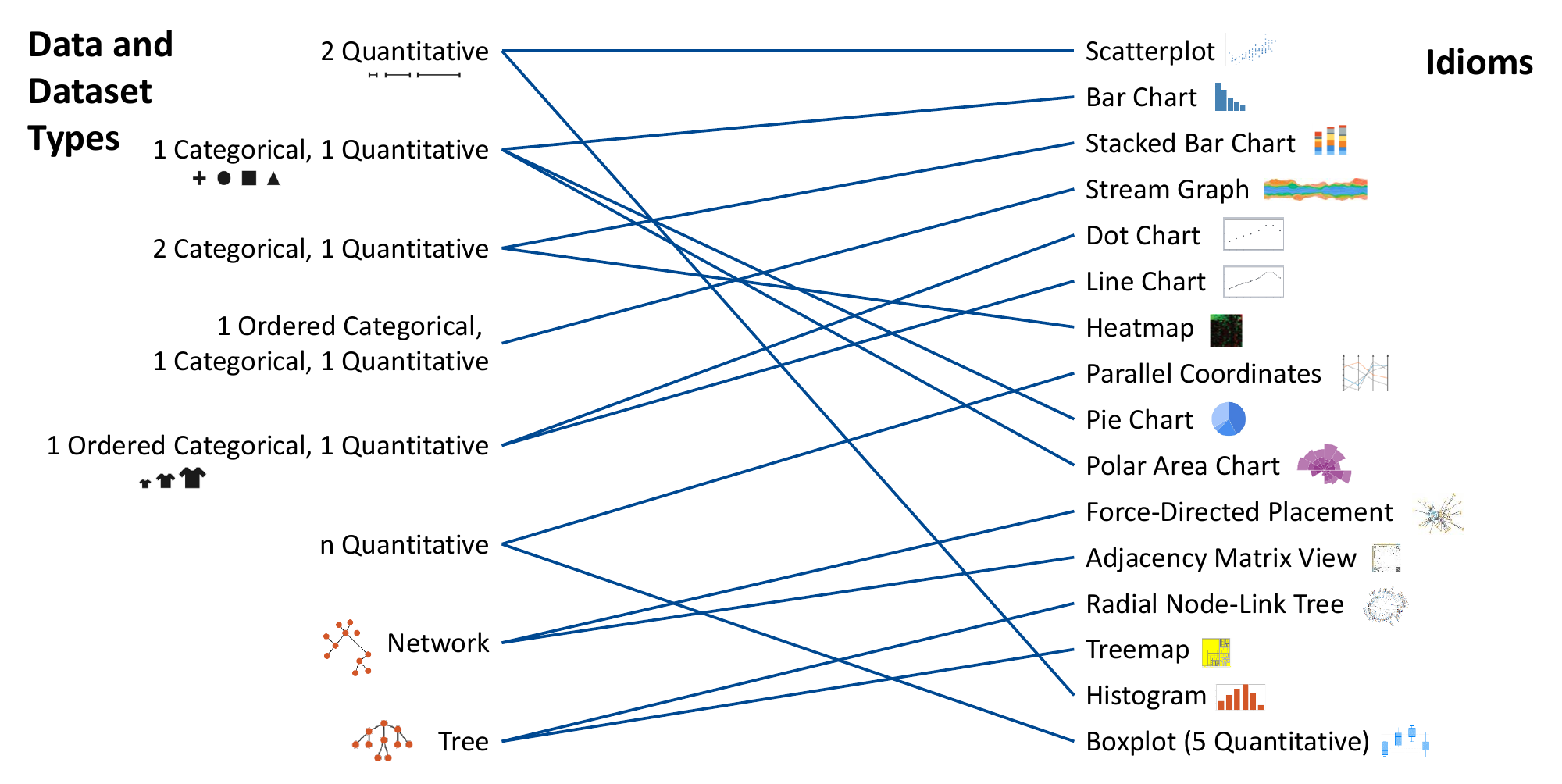}
            \caption{What? $\to$ How?}
            \label{subfig:infovis-mapping-b}
        \end{subfigure}
    \end{center}
    \caption{InfoVis design guidelines mapping \cite{chatti2021designing}}
    \label{fig:infovis}
\end{figure*}

The HCID approach combines Norman's human-centered design (HCD) process \cite{norman2013design} with Munzner's what-why-how visualization framework \cite{munzner2014visualization}, offering a theory-informed method to design LA indicators systematically. The HCID process is composed of four iterative stages that support the design of the appropriate indicator: (1) Define Goal/Question, (2) Ideate, (3) Prototype, and (4) Test. The authors proposed using Indicator Specification Cards (ISCs) to actively engage users in the \textit{Ideate} stage of the HCID process. An ISC aims to allow quick and low-cost design of low-fidelity LA indicators. It follows the Goal-Question-Indicator (GQI) approach \cite{muslim2017goal} to design LA indicators that meet users' goals and applies information visualization (InfoVis) guidelines from Munzner's what-why-how visualization framework \cite{munzner2014visualization}. Concretely, it describes a systematic workflow from the \textit{why?} (i.e., user goal/question) to the \textit{how?} (i.e., visualization). It consists of two main parts, namely \textit{Goal/Question} and \textit{Indicator}, as shown in Figure \ref{fig:isc-example}. The \textit{Goal/Question} part refers to the outcomes of the \textit{Define Goal/Question} stage of the HCID approach. The \textit{Indicator} part is further broken down into three sub-parts, namely \textit{Task Abstraction (Why?)}, \textit{Data Abstraction (What?)}, and \textit{Idiom (How?)}, which reflect the three dimensions of Munzner's what-why-how visualization framework. The InfoVis literature suggests that the \textit{Idioms (How?)} depend heavily on the underlying \textit{Tasks (Why?)} and \textit{Data (What?)} of the visualization and provides guidelines to ``what kind of idioms is more effective for what kind of tasks (mapping Why? $\to$ How?)'' (Figure \ref{subfig:infovis-mapping-a}) and ``what kind of idioms are more effective for what kind of data (mapping What? $\to$ How?)'' (Figure \ref{subfig:infovis-mapping-b}). ISCs were used in \cite{chatti2021designing} to co-design LA indicators to support self-regulated learning (SRL) activities of bachelor students attending an introductory Python programming course. The authors experienced that, while ISCs were initially proposed to support the quick and low-cost design of LA indicators, using ISCs in practice was a complex, time-consuming, and resource-intensive task. To address these limitations, we present the systematic design, implementation, and evaluation details of the \textit{ISC Creator}, an LA tool that aims to enhance the user experience and usefulness of the ISC-based indicator design process.
% ------- In this work, we focus on using and evaluating the existing ISC approach
% ---While HCID primarily aims to give LA stakeholders control over the indicator design and evaluation process to achieve their goals, our work extends user involvement from the design to the development level of the LA process.
% Building upon the idea of 
%%%%%%%%%%%%%%%%%%%%%%%%%%%%%%%%%%%%%%
%%%%%%%%%%%%%%%%%%%%%%%%%%%%%%%%%%%%%%
%%%%%%%%%%%%%%%%%%%%%%%%%%%%%%%%%%%%%%
%%%%%%%%%%%%%%%%%%%%%%%%%%%%%%%%%%%%%%
%%%%%%%%%%%%%%%%%%%%%%%%%%%%%%%%%%%%%%
% \begin{figure}[htpb]
%     \begin{center}
%         \begin{subfigure}[normla]{0.46\textwidth}
%             \centering
%             \includegraphics[width=1\textwidth]{chapters/2-background-related-work/images/infovis-tasks-to-idioms.pdf}
%             \caption{Why? $\to$ How? }
%             \label{subfig:infovis-mapping-a}
%         \end{subfigure}
%         \\
%         \begin{subfigure}[normla]{0.46\textwidth}
%             \centering
%             \includegraphics[width=1\textwidth]{chapters/2-background-related-work/images/infovis-data-to-idioms.pdf}
%             \caption{What? $\to$ How?}
%             \label{subfig:infovis-mapping-b}
%         \end{subfigure}
%     \end{center}
%     \caption{InfoVis design guidelines mapping \cite{chatti2021designing}}
%     \label{fig:specify-goal-question-path}
% \end{figure}
%%%%%%%%%%%%%%%%%%%%%%%%%%%%%%%%%%%%%%
%%%%%%%%%%%%%%%%%%%%%%%%%%%%%%%%%%%%%%%
\section{Human-Centered Design of the ISC Creator}
\label{hcd-isc}
%%%%%%%%%%%%%%%%%%%%%%%%%%%%%%%%%%%%%%%checked
We followed an iterative approach based on the human-centered design (HCD) methodology to understand better the challenges and needs of LA stakeholders, such as teachers and learners, when designing LA indicators. This approach involved consulting with users at each stage of the process to ensure that their needs and requirements were considered. The HCD process consists of four key activities: \textit{Observation}, \textit{Ideation}, \textit{Prototyping}, and \textit{Testing}, which are repeated to gather more insights and get closer to the desired solution with each cycle \cite{norman2013design}.
% Through this initial step, we aimed to understand users' needs and requirements regarding using ISCs to co-design LA indicators.

%%%%%%%%%%%%%%%%%%%%%%%%%%%%%%%%%%%%%%%
\subsection{Observation}
%%%%%%%%%%%%%%%%%%%%%%%%%%%%%%%%%%%%%%%
The primary objective of this initial step was to gain insight into the needs and expectations of users regarding the use of ISCs for co-designing LA indicators. We conducted qualitative user study interviews involving 16 participants to identify the requirements for the \textit{ISC Creator}. The participants included four teachers, seven Bachelor's students, and seven Master's students with diverse educational backgrounds, such as Engineering (Electrical, Embedded Systems), Business Intelligence, and Computer Science. Nine female and seven male participants were from various regions (eight from Asia, six from Europe, and two from North America). Through the interviews, we aimed to accomplish the following objectives: (1) gather users' feedback on their experience with the current ISC, (2) gain a deeper understanding of their thought processes concerning the design of LA indicators, and (3) determine their expectations from using an LA tool to design indicators systematically. We started by showing participants an example ISC (Figure \ref{fig:isc-example}) along with the mappings Why? $\to$ How? and What? $\to$ How?, as provided in Figure \ref{subfig:infovis-mapping-a} and Figure \ref{subfig:infovis-mapping-b}. We explained the purpose of the ISC and the current workflow (i.e., \textit{Task Abstraction}$\to$ \textit{Data Abstraction} $\to$ \textit{Idiom/Chart}) to design LA indicators systematically. We then asked participants to give feedback on the strengths and limitations of the current ISC approach.
% The primary objective of this initial step was to gain insight into the needs and expectations of users regarding the use of ISCs for co-designing LA indicators. We conducted a qualitative user study involving n=16 participants to identify the requirements for the \textit{ISC Creator}. The participants included four teaching assistants/Postdocs and 12 Bachelor's, Master's, and PhD students with diverse educational backgrounds, such as Engineering, Business Intelligence, and Computer Science. Through the interviews, we aimed to accomplish the following objectives: (1) gather users' feedback on their experience with the current ISCs, (2) gain a deeper understanding of their thought processes concerning the design of LA indicators, and (3) determine their expectations from using an LA tool to design indicators systematically. We started by showing participants an example ISC (Figure \ref{fig:isc-example}) along with the mappings Why? $\to$ How? and What? $\to$ How?, as provided in Figure \ref{subfig:infovis-mapping-a} and Figure \ref{subfig:infovis-mapping-b}. We explained the purpose of the ISC and the current workflow (i.e., \textit{Task Abstraction}$\to$ \textit{Data Abstraction} $\to$ \textit{Idiom/Chart}) to design LA indicators systematically. We then asked participants to give feedback on the strengths and limitations of the current ISCs. 
In general, participants expressed a positive opinion about the primary purpose of ISCs. However, they found using the current ISCs to be complex and not straightforward. Most participants (n=13) were concerned that the ISC enforces a sequence of steps for designing indicators. Additionally, they pointed out that selecting appropriate idioms/charts might be confusing for users who lack expertise in InfoVis design practices. At the same time, while participants found the mappings provided in Figure \ref{subfig:infovis-mapping-a} and Figure \ref{subfig:infovis-mapping-b} to be helpful, they still found it challenging to select the appropriate idioms/charts for the given task and data. In addition, many participants (n=10) expressed that the ISC is static and suggested providing interactive ISCs, enabling them to create and customize their indicators based on their specific needs. Additionally, all participants found the information provided in the ISC intangible and desired to see concrete visualizations based on some data.
%%%%%%%%%%%%%%%%%%%%%%%%%%%%%%%%%%%%%%%
\begin{figure}[!ht]
	\centering
	\includegraphics[width=0.49\textwidth]{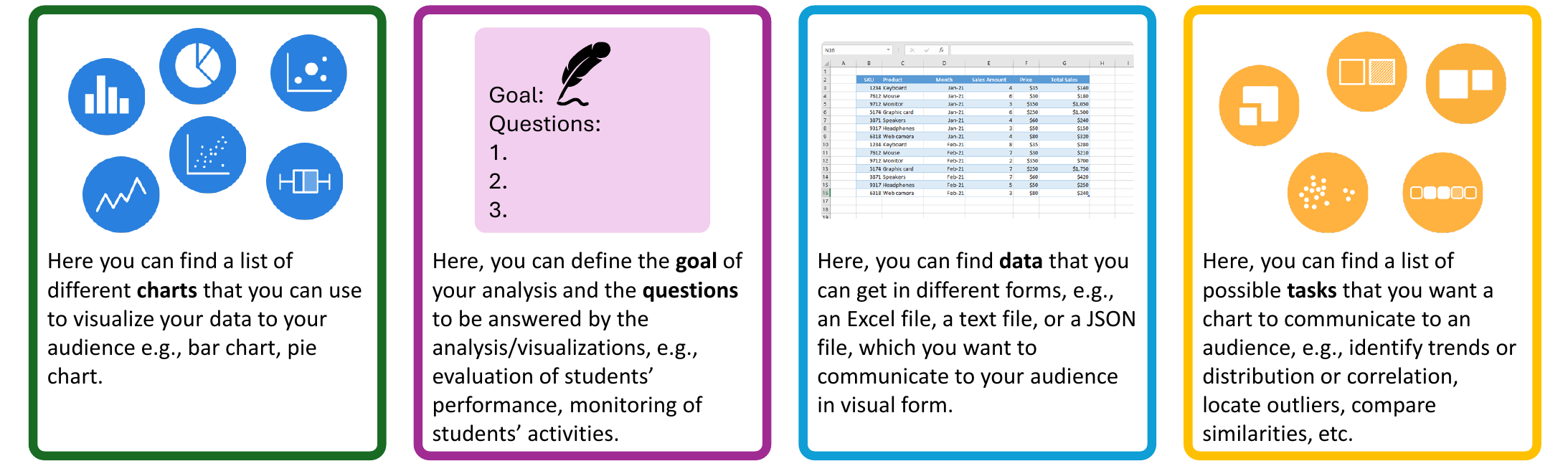}
	\caption{Observation - Cards describing the entry points}
	\label{fig:observation-cards}
\end{figure}

\begin{figure}[!ht]
	\centering
	\includegraphics[width=0.49\textwidth]{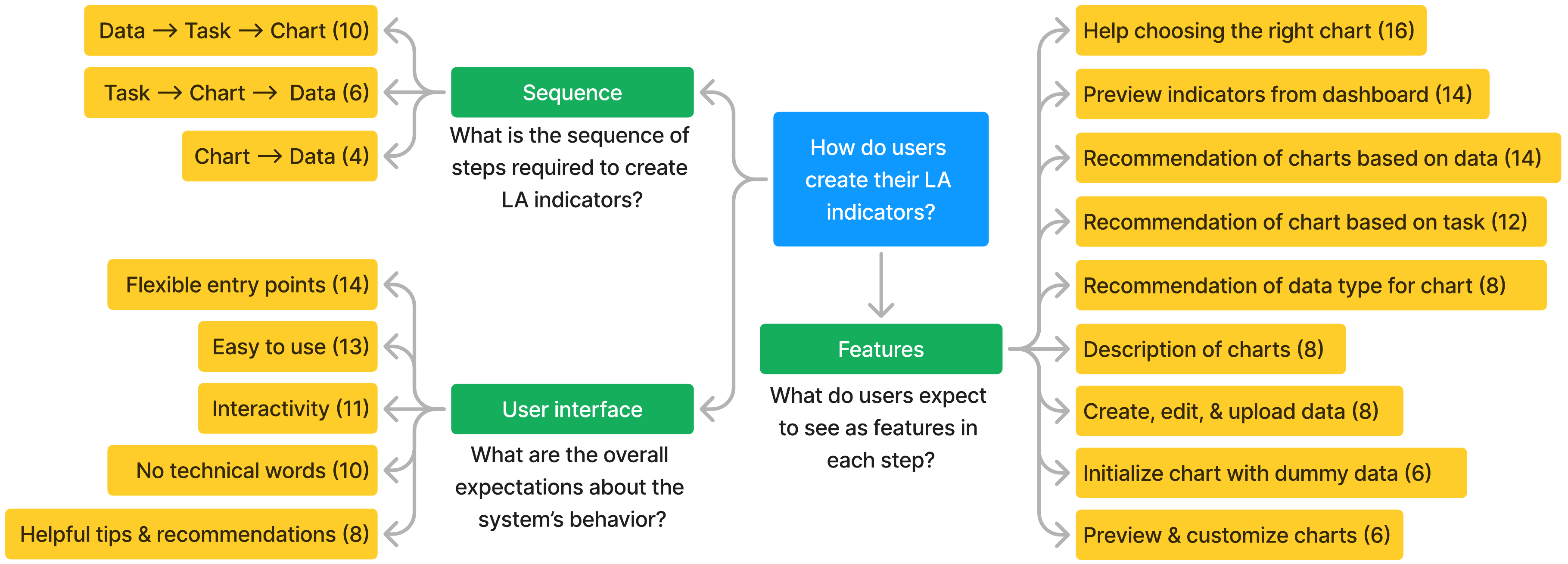}
	\caption{Observation - Thematic Analysis}
	\label{fig:thematic-analysis}
\end{figure}

%%%%%%%%%%%%%%%%%%%%%%%%%%%%%%%%%%%%%%%
To better understand how users would design LA indicators, we further presented participants with four cards illustrating the necessary steps and different entry points for creating LA indicators, as illustrated in Figure \ref{fig:observation-cards}. We then asked them to describe the sequence of cards they preferred and the features they expected at each step. The interviews, lasting 30 to 45 minutes, were recorded and transcribed with the participant's consent. After collecting notes, transcripts, and recordings of the interviews, we conducted an iterative thematic analysis based on the guidelines provided by Braun and Clark \cite{braun2006using}. We familiarized ourselves with the data and systematically coded the transcripts. We then organized these codes into coherent themes. Our inductive approach helped us identify three major themes in response to how users prefer to create LA indicators: \textit{User Interface} (overall expectations about the system's behavior), \textit{Sequence} (the required steps for creating LA indicators), and \textit{Content/Features} (expected features in each step). Figure \ref{fig:thematic-analysis} illustrates the thematic analysis results, with themes highlighted in green and codes in yellow. Each code is accompanied by the number of times participants mentioned it.
%%%%%%%%%%%%%%%%%%%%%%%%%%%%%%%%%%%%%%%

%%%%%%%%%%%%%%%%%%%%%%%%%%%%%%%%%%%%%%%
From our observations, we developed a set of design goals (DG) to guide our functional \textit{ISC Creator} prototypes:

\textbf{(DG1) Intuitive User Interface:} Participants expressed the need for an interactive and intuitive User Interface (UI) for the \textit{ISC Creator} to simplify designing LA indicators. They explained that they found the current ISC difficult to understand, especially the technical terms like ``How to encode?'' and ``Derived data.'' Additionally, they emphasized the importance of enhancing the static ISCs by enabling interaction with the \textit{ISC Creator}. This could be achieved by providing an overview of created ISCs in a dashboard, allowing users to see details as needed, and navigating between different UI panels to select, preview, and edit charts and data. They suggested including features like chart recommendations, examples, and descriptions and simple methods for customizing charts and populating data.

\textbf{(DG2) Flexible sequence of steps:} After gathering feedback from the participants, it was found that the predetermined sequence of steps in the current ISC (\textit{Task Abstraction} $\to$ \textit{Data Abstraction} $\to$ \textit{Idiom/Chart}) was considered to be too restrictive. We found that participants had varying opinions on following a specific sequence for designing LA indicators. Therefore, the system should allow users to define the order of steps for designing indicators based on their requirements.

\textbf{(DG3) Recommendations of idioms/charts:} To ensure practical usage of the \textit{ISC Creator}, the system should include recommendations for suitable idioms/charts based on the user's selected data and task, following the guidelines from the InfoVis literature. Additionally, the system should assist users in choosing the appropriate data to populate the chart for optimal results.
%%%%%%%%%%%%%%%%%%%%%%%%%%%%%%%%%%%%%%%
\subsection{Ideation}
\begin{figure}[!ht]
	\centering
	\includegraphics[width=0.49\textwidth]{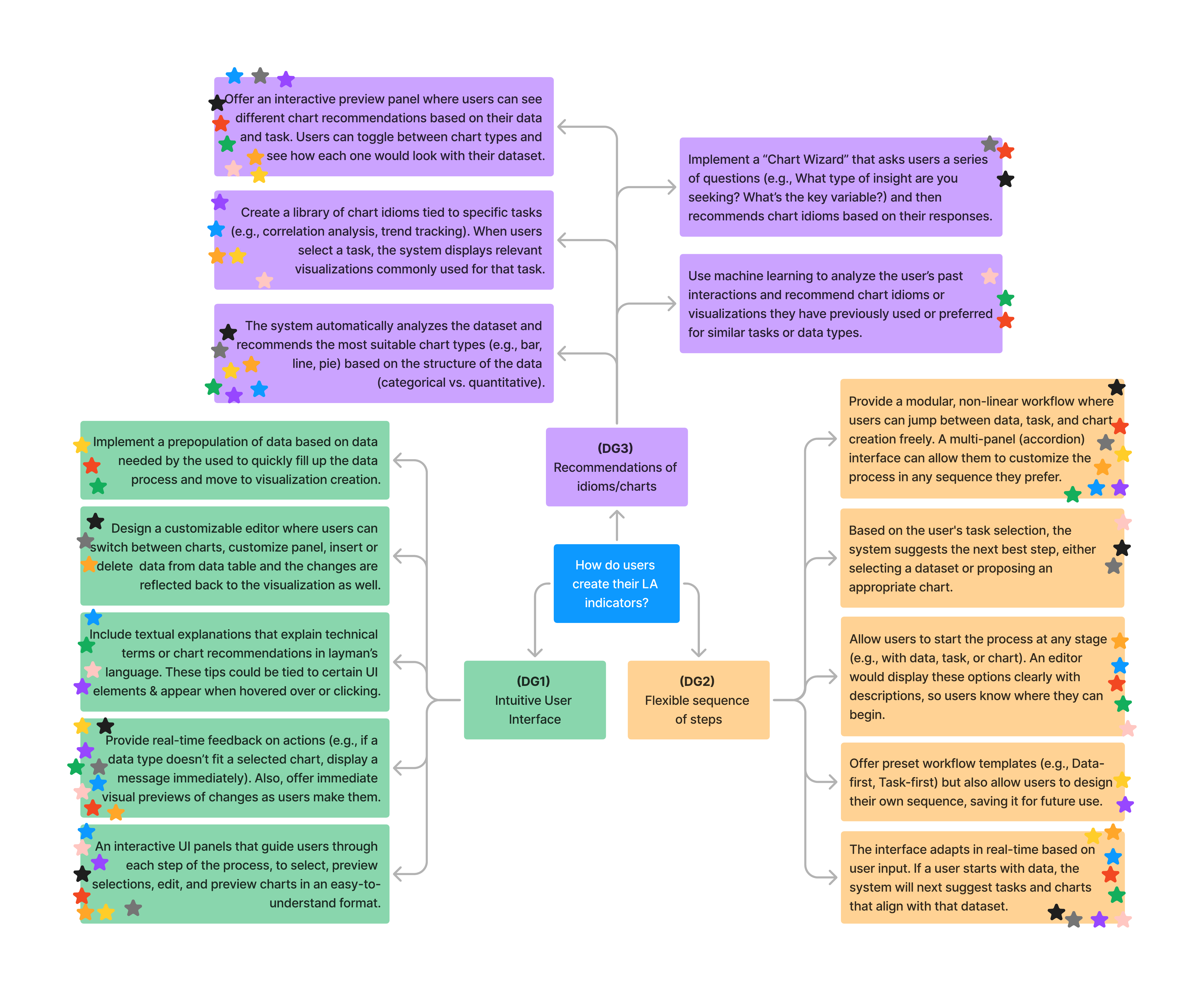}\\
	\caption{Ideation - Brainstorming Session}
	\label{fig:ideation}
\end{figure}

%%%%%%%%%%%%%%%%%%%%%%%%%%%%%%%%%%%%%%%
We conducted an online brainstorming session with four PhD students and five Master's students from the local university, all with strong knowledge of LA, data analytics, and InfoVis. The goal was to gather as many ideas as possible for each DG based on user requirements identified during the observation phase, emphasizing quantity over quality. For each DG, every idea was recorded on a sticky note. Following a ``pitch and critique'' approach, participants pitched their ideas to the group while receiving positive and negative feedback. 
% The final step was a voting process to select the top ideas. Each participant was allotted three votes to choose the three best ideas. Figure \ref{fig:ideation} shows various ideas for each DG marked with star ratings.
During the pitch, each participant presented their ideas to the group, explaining their concept and how it addressed the related DG. The critique part followed, where team members provided feedback by highlighting the strengths and weaknesses of each idea, always aiming to refine them for a better user-centered approach. This interactive process encouraged generating new ideas and iterative improvements through collaborative discussion. The final step involved a voting process to select the top ideas. Each participant was allotted three votes to choose the three best ideas, marking their favorite ideas with star ratings. Figure \ref{fig:ideation} illustrates various summarized ideas for each DG, where participants voted with stars to signify the most favored ideas.
%%%%%%%%%%%%%%%%%%%%%%%%%%%%%%%%%%%%%%%

%%%%%%%%%%%%%%%%%%%%%%%%%%%%%%%%%%%%%%%
One of the DGs was developing an intuitive UI \textbf{(DG1)}. Therefore, the UI must focus on enhancing the user experience, which includes prepopulating data based on user needs to speed up the input process and providing a customizable editor where users can seamlessly switch between charts, modify selections, and update data with real-time reflections in the visualization. Textual explanations in plain language must clarify technical terms and chart recommendations, appearing contextually when hovered over or clicked. Real-time feedback alerts users to errors or incompatible data types while providing instant visual previews of changes. An interactive UI panel guides users step-by-step, simplifying the entire data selection process, chart editing, and visualization creation.
% the most popular suggestions were to have a separate interface for viewing the list of users' ISCs in a dashboard and another interface to create/edit ISCs. The interface of the \textit{ISC Creator} must be user-friendly, with distinct panels that segregate the sections for selecting charts and data and provide users with easy manipulation. In addition, examples and user-friendly labels should be provided to help users effectively utilize the tool.

As part of achieving the DG of creating a flexible sequence of steps to design LA indicators \textbf{(DG2)}, 
the design must emphasize user control and adaptability. It must offer a modular, non-linear workflow through a multi-panel accordion interface, allowing users to freely jump between data, task, and chart creation in any preferred sequence. The interface should adapt in real-time based on user input, suggesting tasks and charts that align with the data when the user starts with data. Additionally, users can begin the process at any stage, with a clear editor displaying options and descriptions to guide them in selecting where to start, whether with data, tasks, or charts.
% the most voted ideas express that users tend to have different entry points and sequences of steps to create their indicators. Concretely, users should be able to move between the selection of tasks, data, and idioms/charts flexibly \textbf{(DG2)}. 
% Moreover, recommendations for appropriate idioms/charts based on the selected data and/or task should be provided \textbf{(DG3)}. 

Lastly, for the DG of recommending idioms/charts \textbf{(DG3)}, the system must focus on providing intelligent, data-driven suggestions. It automatically understands the dataset and recommends suitable chart types (e.g., bar, line, pie) based on the data structure, such as categorical or numerical data types. An interactive preview panel must allow users to toggle between different chart recommendations and instantly see how each would look with their data. Relevant visualizations must be filtered accordingly based on the selected task.
%%%%%%%%%%%%%%%%%%%%%%%%%%%%%%%%%%%%%%%
\subsection{Prototyping and Testing}
\begin{figure}[!ht]
	\centering
	\includegraphics[width=0.49\textwidth]{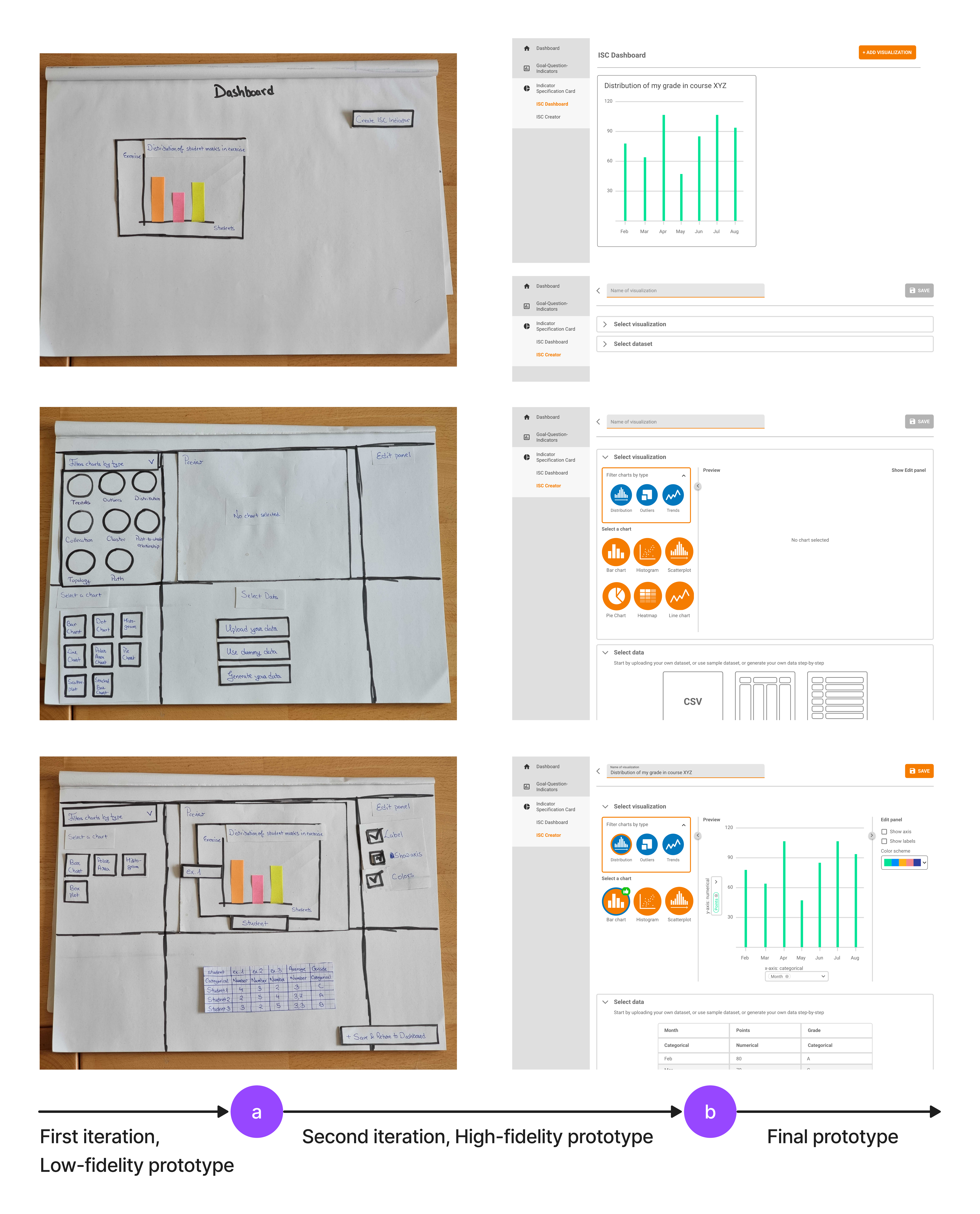}\\
	\caption{The two iterations of the human-centered design (a) low-fidelity prototypes and (b) high-fidelity prototypes}
	\label{fig:prototyping-testing}
\end{figure}

%%%%%%%%%%%%%%%%%%%%%%%%%%%%%%%%%%%%%%%
Following the observation and ideation phases, we initiated the prototyping phase, which began with low-fidelity prototypes and progressed to high-fidelity prototypes. Throughout all iterations, we maintained close communication with end-users. In each iteration, we recruited a minimum of five new users to provide feedback, and their feedback was incorporated into the subsequent design. Figure \ref{fig:prototyping-testing} presents an overview of the iterative design process.
%%%%%%%%%%%%%%%%%%%%%%%%%%%%%%%%%%%%%%%

%%%%%%%%%%%%%%%%%%%%%%%%%%%%%%%%%%%%%%%
%%%%%%%%%%%%%%%%%%%%%%%%%%%%%%%%%%%%%%%
\subsubsection{First iteration, initial design, low-fidelity prototypes} 
%%%%%%%%%%%%%%%%%%%%%%%%%%%%%%%%%%%%%%%
The initial interactive prototypes were developed to gather user feedback on the \textit{ISC Creator}, considering the users' needs identified during the observation phase. Based on the results from the ideation phase, we created a preliminary design using low-fidelity prototypes, starting with paper prototypes due to their simplicity and convenience. The first step involved designing a basic dashboard to allow users to preview saved ISCs and create new ones \textbf{(DG1)}. Users can begin the indicator design process by selecting a chart, filtering charts by type or task, or choosing data \textbf{(DG1, DG2)}. The interface is designed always to display both the charts and the data table, enabling easy switching between steps \textbf{(DG1, DG2)}. The left panel shows a list of charts that are compatible with the selected task and/or data \textbf{(DG1, DG3)}, while the right panel provides an editing section where users can customize the chart, such as by adding labels \textbf{(DG1)}.

To evaluate the low-fidelity prototypes, we conducted interview sessions with six participants (four males and two females), including two Bachelor's students, two Master's students, and two PhD students. Most participants (n=4) were unfamiliar with LA concepts and data analytics. Before the interviews, we prepared a list of core activities for the participants to design an LA indicator using the \textit{ISC Creator}, such as selecting a chart, uploading a dataset, using filters, customizing the chart, and adding the chart to the dashboard. Overall, participants found the \textit{ISC Creator}'s UI elements clear and the interaction to select charts, tasks, and data straightforward. Moreover, they expressed positive opinions towards the flexibility of switching between steps. However, they found the amount of information provided on the UI overwhelming, and many were confused about where to begin. Concretely, they wanted clear entry points to help them select the appropriate sequence of steps to design their indicators. We used this feedback in the next design iteration to enhance the low-fidelity prototypes.
%%%%%%%%%%%%%%%%%%%%%%%%%%%%%%%%%%%%%%%
\subsubsection{Second iteration, high-fidelity prototypes} 
%%%%%%%%%%%%%%%%%%%%%%%%%%%%%%%%%%%%%%%
In this iteration, we created high-fidelity prototypes using Figma. One of the key issues mentioned by the users from the previous iteration was that they needed a clear entry point. Therefore, we introduced three flexible approaches to design indicators \textbf{(DG2)}, namely: (1) \textit{task-driven approach}, (2) \textit{data-driven approach}, and (3) \textit{visualization-driven approach}. For all the approaches, we integrated recommendations of suitable idioms/charts based on chosen data and/or task at hand \textbf{(DG3)}.
\begin{itemize}
    \item \textbf{Task-driven approach:} Users have a specific goal in mind, and based on their goal, they can choose an appropriate task, then choose a recommended idiom/chart, and finally select the appropriate data (i.e., \textit{Task Abstraction} $\to$ \textit{Idiom/Chart} $\to$ \textit{Data Abstraction}).
    \item \textbf{Data-driven approach:} Users have data and based on the data types, users can either choose a recommended idiom/chart directly (i.e., \textit{Data Abstraction} $\to$ \textit{Idiom/Chart}) or choose a task and then choose a recommended idiom/chart (i.e., \textit{Data Abstraction} $\to$ \textit{Task Abstraction} $\to$ \textit{Idiom/Chart}).
    \item \textbf{Visualization-driven approach:} Users can choose their desired idiom/chart directly, without having to choose the task, and then fill the selected chart with data (i.e., \textit{Idiom/Chart} $\to$ \textit{Data Abstraction})
\end{itemize}
% The \textit{task-driven approach} allows users to select a specific task based on their goal, choose a recommended idiom/chart, and select the appropriate data. The sequence for this approach is \textit{Task Abstraction} $\to$ \textit{Idiom/Chart} $\to$ \textit{Data Abstraction}. 
% The \textit{data-driven approach} gives users the ability to work with their data, where they can either directly choose a recommended idiom/chart based on the data types (\textit{Data Abstraction} $\to$ \textit{Idiom/Chart}) or choose a task and then select a recommended idiom/chart (\textit{Data Abstraction} $\to$ \textit{Task Abstraction} $\to$ \textit{Idiom/Chart}). 
% Lastly, the \textit{visualization-driven approach} lets users choose their desired idiom/chart directly and then fill it with data. The sequence for this approach is \textit{Idiom/Chart} $\to$ \textit{Data Abstraction}. Users now have the flexibility to start with the visualization or the data \textbf{(DG2)}. Users can filter charts by intended task \textbf{(DG2)}. Furthermore, we use thumbs-up icons to recommend appropriate idioms/charts based on the selected data and/or task \textbf{(DG1, DG3)}.

We gathered feedback on the high-fidelity prototypes from seven participants (five Bachelor's students and two teaching assistants). All participants were unfamiliar with LA concepts and data analytics. Two participants had participated in the previous evaluation, while the other five were new to the study. We asked them to complete a task that involved designing an indicator to monitor students' performance distribution in their course. Throughout the task, we encouraged them to think aloud. Overall, the participants responded positively to the prototypes and provided constructive criticism and suggestions for improvement. Specifically, they recommended improving the data editing functionality, displaying the column data type and changing data types in the ``Select Data'' section, and making the selected chart more prominent on the canvas, such as spanning the entire width or displaying it in full-screen.
Additionally, three participants expressed confusion with the recommendations for specific charts. They requested further explanation regarding whether the recommendation was based on the selected task or the data types in the data table. They also suggested adding the functionality to delete, edit, and duplicate the chart from the dashboard and the option to download the indicator as an image.
A notable finding from the interview was that many participants (n=5) initially struggled to concretely understand and formulate the goal of the indicator and identify the appropriate data supporting it. This difficulty highlighted the need for more precise guidance and support within the prototypes to help users better define their objectives and define the data and its type.
%%%%%%%%%%%%%%%%%%%%%%%%%%%%%%%%%%%%%%%
\subsection{Final Prototypes and Implementation}
\begin{figure}[htpb]
    \begin{center}
        \begin{subfigure}[normla]{0.23\textwidth}
            \includegraphics[width=1\textwidth]{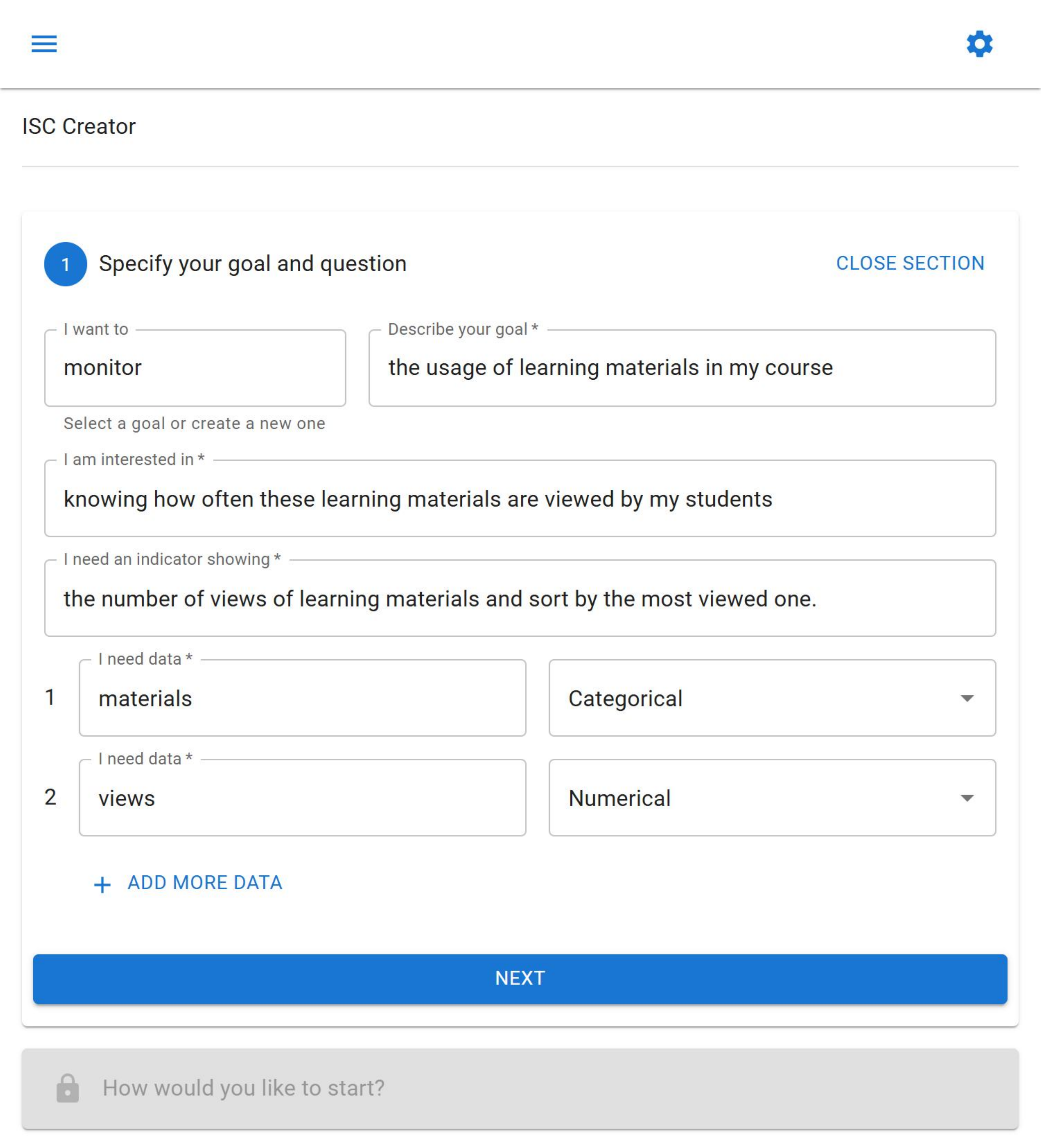}
            \caption{Specify a goal and formulate a question}
            \label{subfig:specify-goal-question}
        \end{subfigure}
        ~
        \begin{subfigure}[normla]{0.23\textwidth}
            \includegraphics[width=1\textwidth]{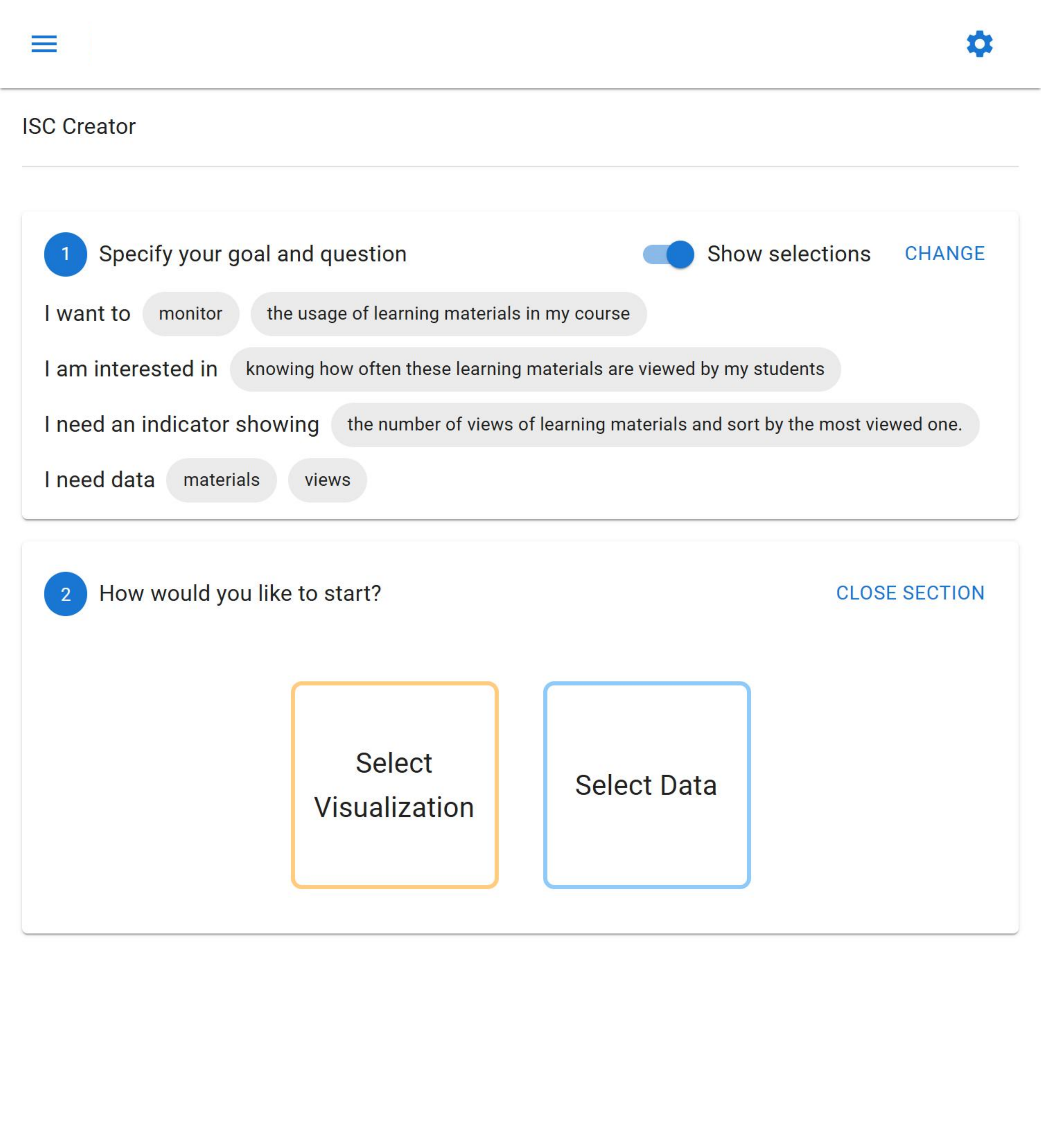}
            \caption{Choose path: Visualization (\textit{Task}|\textit{Idiom}) or Dataset (\textit{Data})}
            \label{subfig:specify-path}
        \end{subfigure}
    \end{center}
    \caption{Specify goal, question, and choose path}
    \label{fig:specify-goal-question-path}
\end{figure}

%%%%%%%%%%%%%%%%%%%%%%%%%%%%%%%%%%%%%%%
After incorporating feedback from previous iterations, we improved the high-fidelity prototypes. We developed the final prototypes of the \textit{ISC Creator} using React.js and Material Design, and for visualizations, we used ApexCharts.js. As shown in Figure \ref{subfig:dashboard}, a user can view a list of their designed indicators as a list in the ISC dashboard and create a new ISC by clicking the `CREATE NEW' button from the dashboard to direct them to the \textit{ISC Creator} page. One of the critical issues from the previous iteration was that the participants had challenges understanding and formulating the goal of the indicator concretely. Therefore, as shown in Figure \ref{subfig:specify-goal-question}, we introduced a new section ``Specify your goal and question'', where a user can concretely specify their goal, their idea for the indicator, and the data required to design the ISC (minimum of two). Users can also determine the data type (categorical, numerical, categorical (ordered)) for each data. Once a user has brainstormed a rough idea of their goal, they can move on to the next step.
As shown in Figure \ref{subfig:specify-path}, the user can start either by selecting visualization (\textit{task-driven approach}/\textit{visualization-driven approach}) or selecting dataset (\textit{data-driven approach}). This allows user to start their interaction from different parts of the \textit{ISC Creator} \textbf{(DG2)}. The final details of the three different approaches to prototyping are discussed in the following sections. 
% TODO: Add details to the GQI
%To illustrate the capabilities of the \textit{ISC Creator}, we presented two examples: (1) a student who desires to monitor the distribution of their grades in a course, and (2) a teacher who aims to monitor the most viewed learning material in their course.
%%%%%%%%%%%%%%%%%%%%%%%%%%%%%%%%%%%%%%%
% Specify goal, question, and path

%%%%%%%%%%%%%%%%%%%%%%%%%%%%%%%%%%%%%%%
%%%%%%%%%%%%%%%%%%%%%%%%%%%%%%%%%%%%%%%
\subsubsection{Task-driven Approach}
\begin{figure}[htpb]
    \begin{center}
        \begin{subfigure}[normla]{0.23\textwidth}
            \includegraphics[width=1\textwidth]{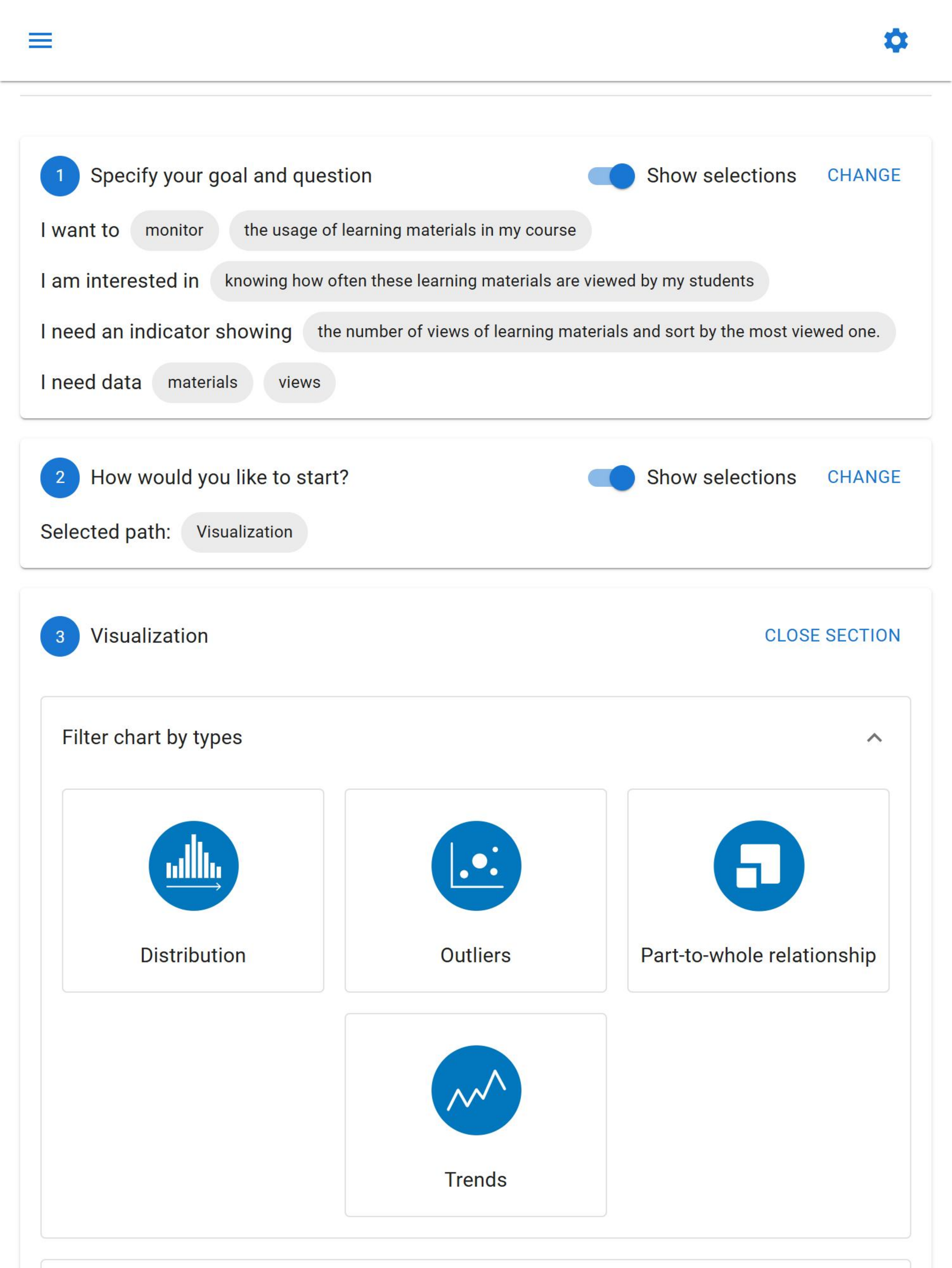}
            \caption{Specify a \textit{Task}}
            \label{subfig:view-tasks}
        \end{subfigure}
        ~
        \begin{subfigure}[normla]{0.23\textwidth}
            \includegraphics[width=1\textwidth]{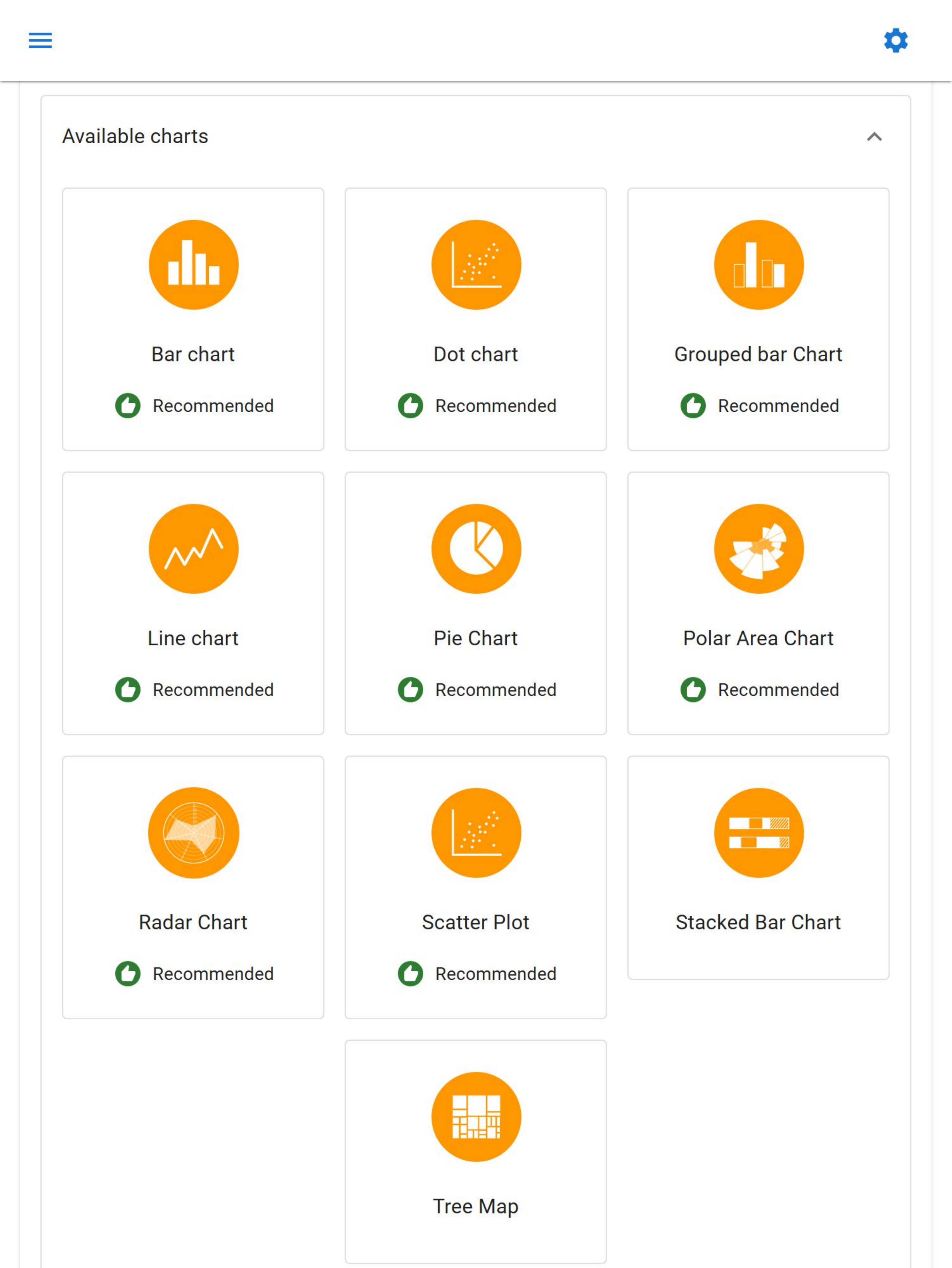}
            \caption{Choose an \textit{Idiom}}
            \label{subfig:view-charts}
        \end{subfigure}
        \\
        \begin{subfigure}[normla]{0.23\textwidth}
            \includegraphics[width=1\textwidth]{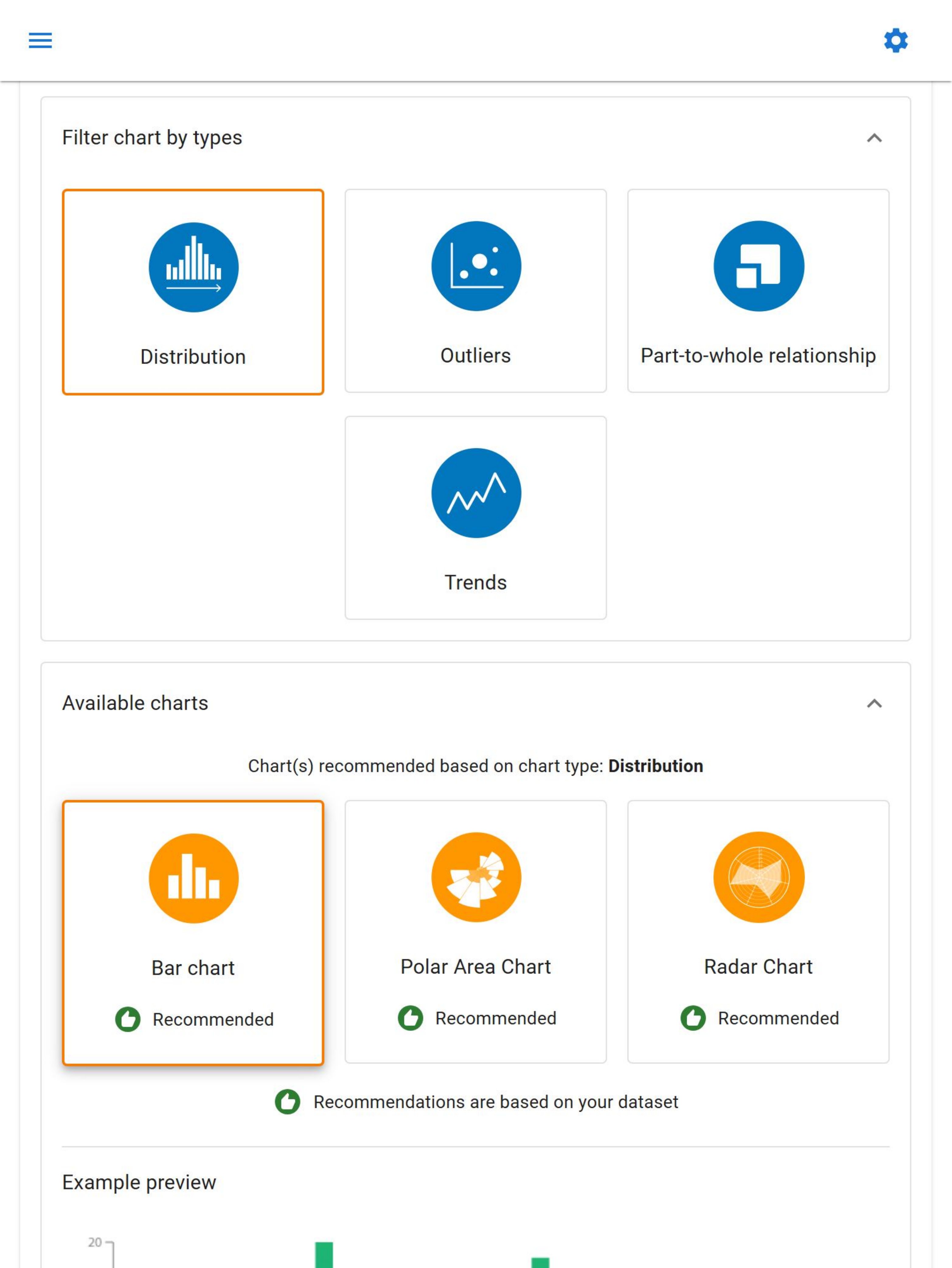}
            \caption{Recommend \textit{Idioms} based on \textit{Task} and \textit{Data}}
            \label{subfig:select-task-chart}
        \end{subfigure}
        ~
        \begin{subfigure}[normla]{0.23\textwidth}
            \includegraphics[width=1\textwidth]{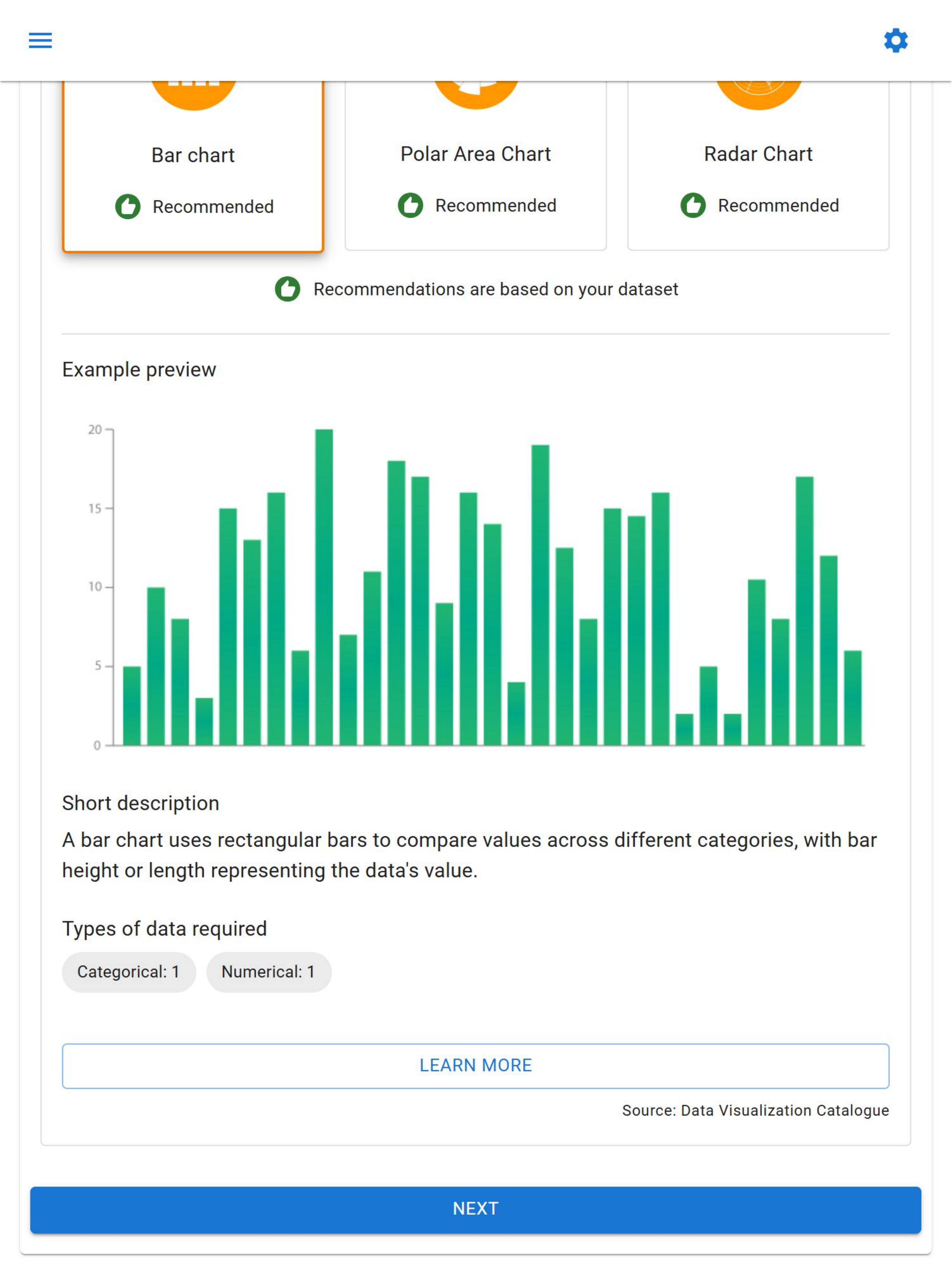}
            \caption{Preview details of selected \textit{Idiom}}
            \label{subfig:selected-chart-details}
        \end{subfigure}
    \end{center}
    \caption{Task/Visualization driven approach}
    \label{fig:task-vis-driven}
\end{figure}

%%%%%%%%%%%%%%%%%%%%%%%%%%%%%%%%%%%%%%%
Suppose the user clicks on \textit{Select Visualization}. In that case, they are presented with a list of tasks (Figure \ref{subfig:view-tasks}) and a list of idioms/charts (Figure \ref{subfig:view-charts}), each accompanied by small illustrations for quick recognition of the chart types \textbf{(DG1)} and descriptions that appear when hovered over these UI elements. The user can select a specific task of their choice, and the \textit{ISC Creator} will recommend a list of idioms/charts that are suitable for both the selected task and data types \textbf{(DG3)}. As shown in Figure \ref{subfig:select-task-chart}, the \textit{ISC Creator} provides an explanation when a task (e.g., \textit{Distribution}) is selected. 
Moreover, thumbs-up icons appear close to the charts, which signifies a recommendation based on the chosen dataset. As shown in Figure \ref{subfig:specify-goal-question}, in the `Specify your goal and question' UI section, the user specified the type of data (i.e., one categorical and one numerical) that matches the requirements for creating a chart, e.g., a bar chart. This explanation is provided under the detailed view of the selected chart when scrolled below \textbf{(DG3)} (Figure \ref{subfig:selected-chart-details}). Next, the user can click the `NEXT' button to select data (Figure \ref{subfig:create-data}) and finalize the visualization (Figure \ref{fig:finalize-indicator-step}).
%%%%%%%%%%%%%%%%%%%%%%%%%%%%%%%%%%%%%%%
% Task/Visualization driven approach

%%%%%%%%%%%%%%%%%%%%%%%%%%%%%%%%%%%%%%%
%%%%%%%%%%%%%%%%%%%%%%%%%%%%%%%%%%%%%%%
\subsubsection{Data-driven Approach}
\begin{figure}[htpb]
    \begin{center}
        \begin{subfigure}[normla]{0.23\textwidth}
            \includegraphics[width=1\textwidth]{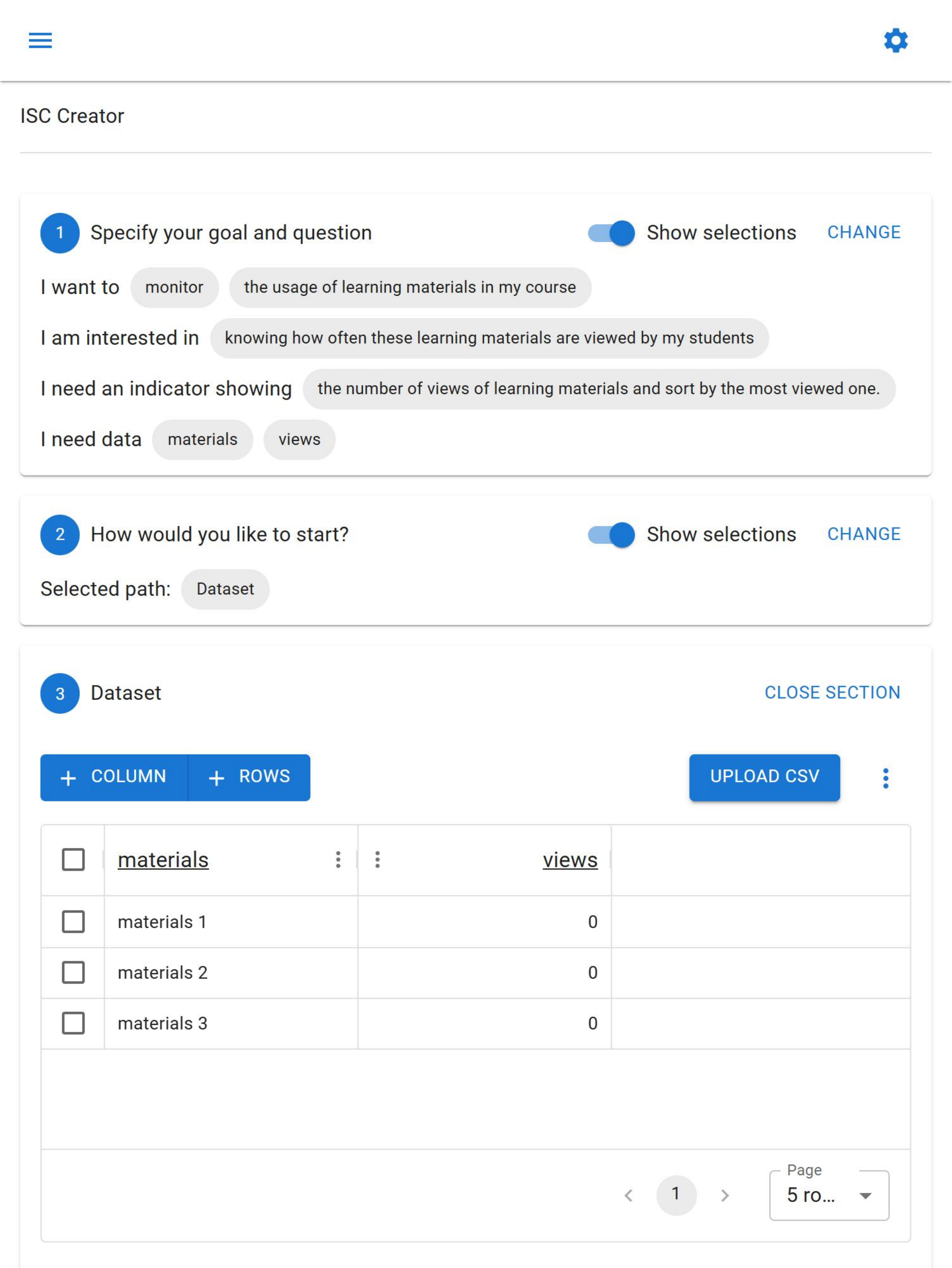}
            \caption{Create own data}
            \label{subfig:create-data}
        \end{subfigure}
        ~
        \begin{subfigure}[normla]{0.23\textwidth}
            \includegraphics[width=1\textwidth]{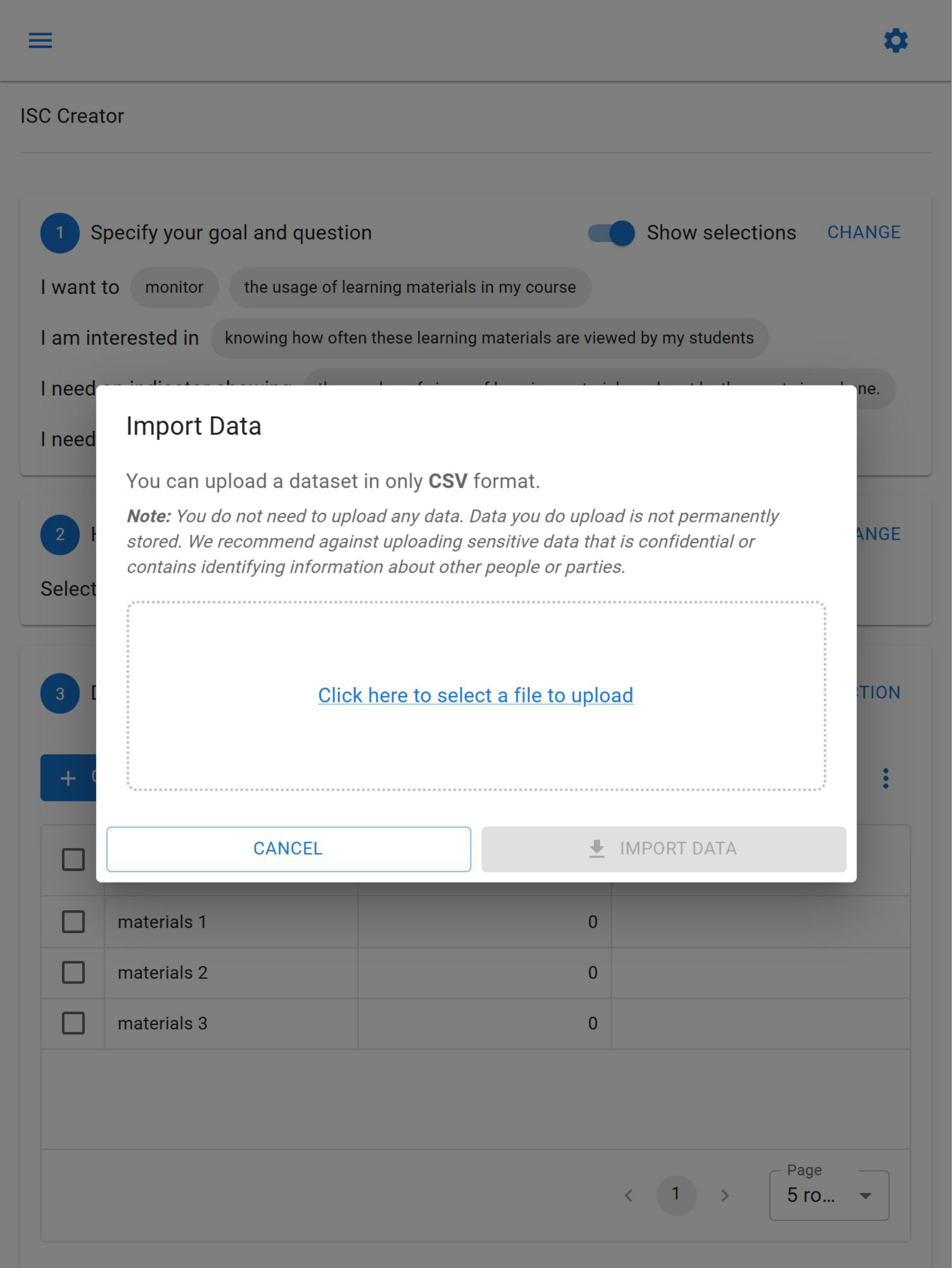}
            \caption{Upload a CSV data}
            \label{subfig:upload-data-1}
        \end{subfigure}
        \\
        \begin{subfigure}[normla]{0.23\textwidth}
            \includegraphics[width=1\textwidth]{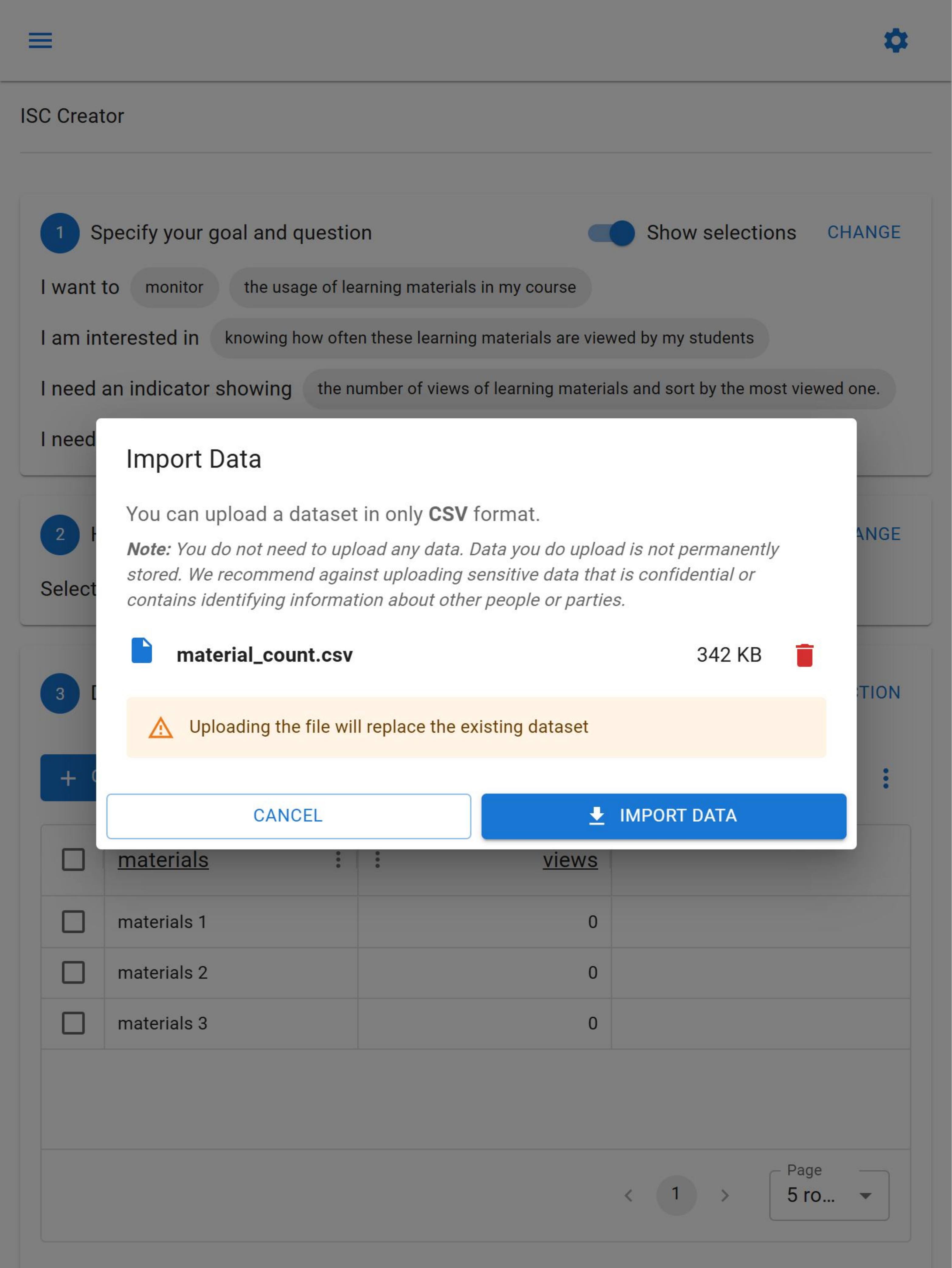}
            \caption{Import a CSV file to table}
            \label{subfig:upload-data-confirm}
        \end{subfigure}
        ~
        \begin{subfigure}[normla]{0.23\textwidth}
            \includegraphics[width=1\textwidth]{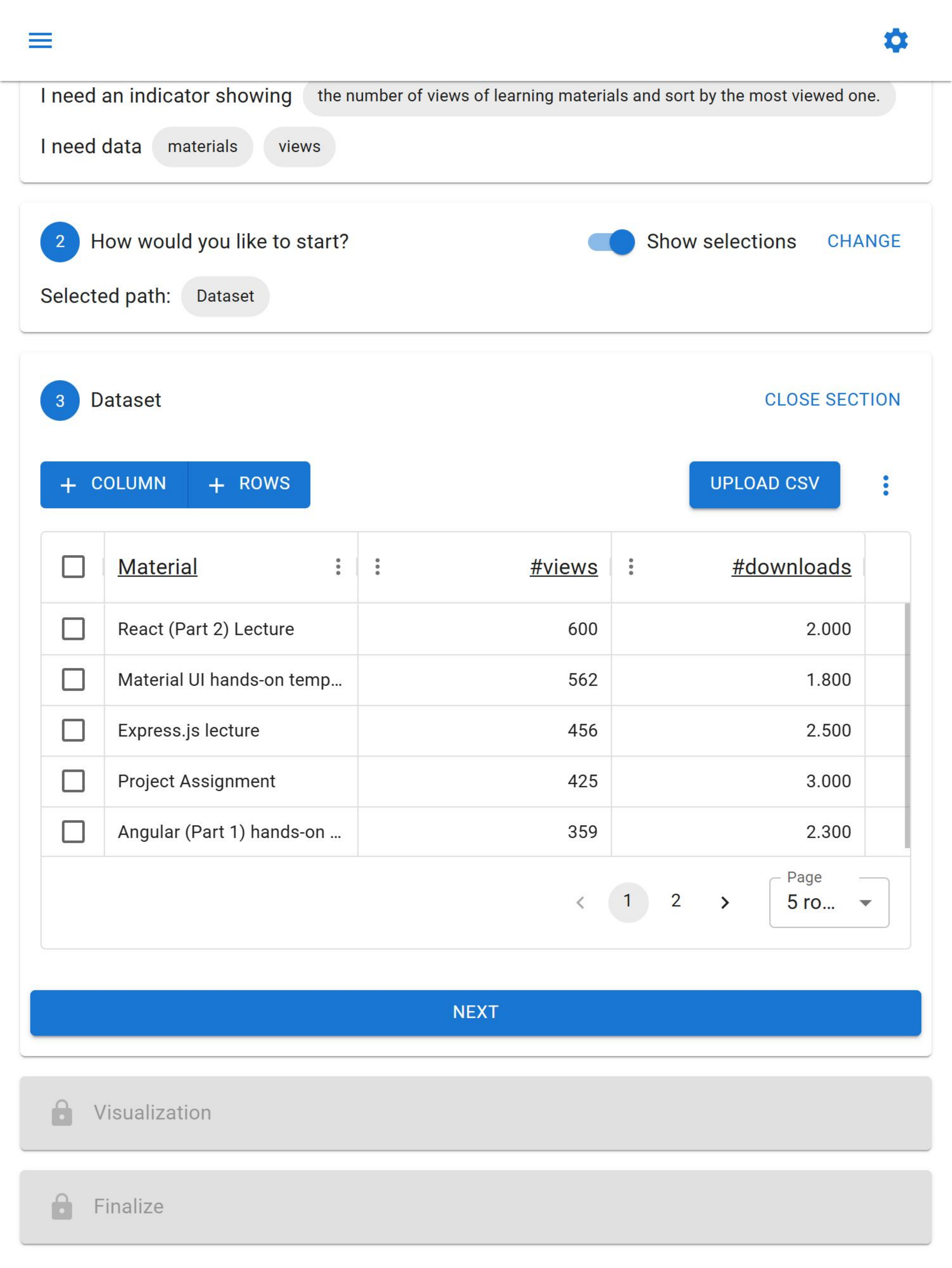}
            \caption{Preview the CSV data}
            \label{subfig:preview-data}
        \end{subfigure}
    \end{center}
    \caption{Data-driven approach}
    \label{fig:data-driven}
\end{figure}

%%%%%%%%%%%%%%%%%%%%%%%%%%%%%%%%%%%%%%%
Suppose the user clicks on \textit{Select Dataset}. In this case, they are presented with a data table (Figure \ref{subfig:create-data}) prepopulated with sample data and column names, which correspond to the data specified in the `Specify your goal and question' UI section (Figure \ref{subfig:specify-goal-question}). As shown in Figure \ref{subfig:create-data}, the user can add/remove columns and rows as needed \textbf{(DG1)}. Additionally, the user can upload a CSV file by clicking the `UPLOAD CSV' button \textbf{(DG1)}, which opens a dialog box to select and upload a file (Figures \ref{subfig:upload-data-1} and \ref{subfig:upload-data-confirm}). Once the user clicks the `IMPORT DATA' button, the data is loaded into the data table (Figure \ref{subfig:preview-data}). The \textit{ISC Creator} automatically detects the data types for each column in the uploaded file \textbf{(DG1)}. In this example, the user uploaded a CSV file that includes one categorical and two numerical data type columns. Users can then click `NEXT' to proceed to select a task (Figure \ref{subfig:view-tasks}) and/or chart (Figure \ref{subfig:view-charts}). When finalizing the visualization, the user is recommended the appropriate columns for each chart axis \textbf{(DG3)}. For instance, as shown in Figure \ref{subfig:selected-chart-details}, a bar chart requires one categorical and one numerical data type. In Figure \ref{subfig:finalize-indicator}, all categorical data type columns from the data table are displayed on the x-axis dropdown, while numerical data type columns are mapped to the y-axis dropdown, respectively.
%%%%%%%%%%%%%%%%%%%%%%%%%%%%%%%%%%%%%%%
% Data-driven

%%%%%%%%%%%%%%%%%%%%%%%%%%%%%%%%%%%%%%%
%%%%%%%%%%%%%%%%%%%%%%%%%%%%%%%%%%%%%%%
\subsubsection{Visualization-driven Approach}
\begin{figure}[htpb]
    \begin{center}
        \begin{subfigure}[normla]{0.23\textwidth}
            \includegraphics[width=1\textwidth]{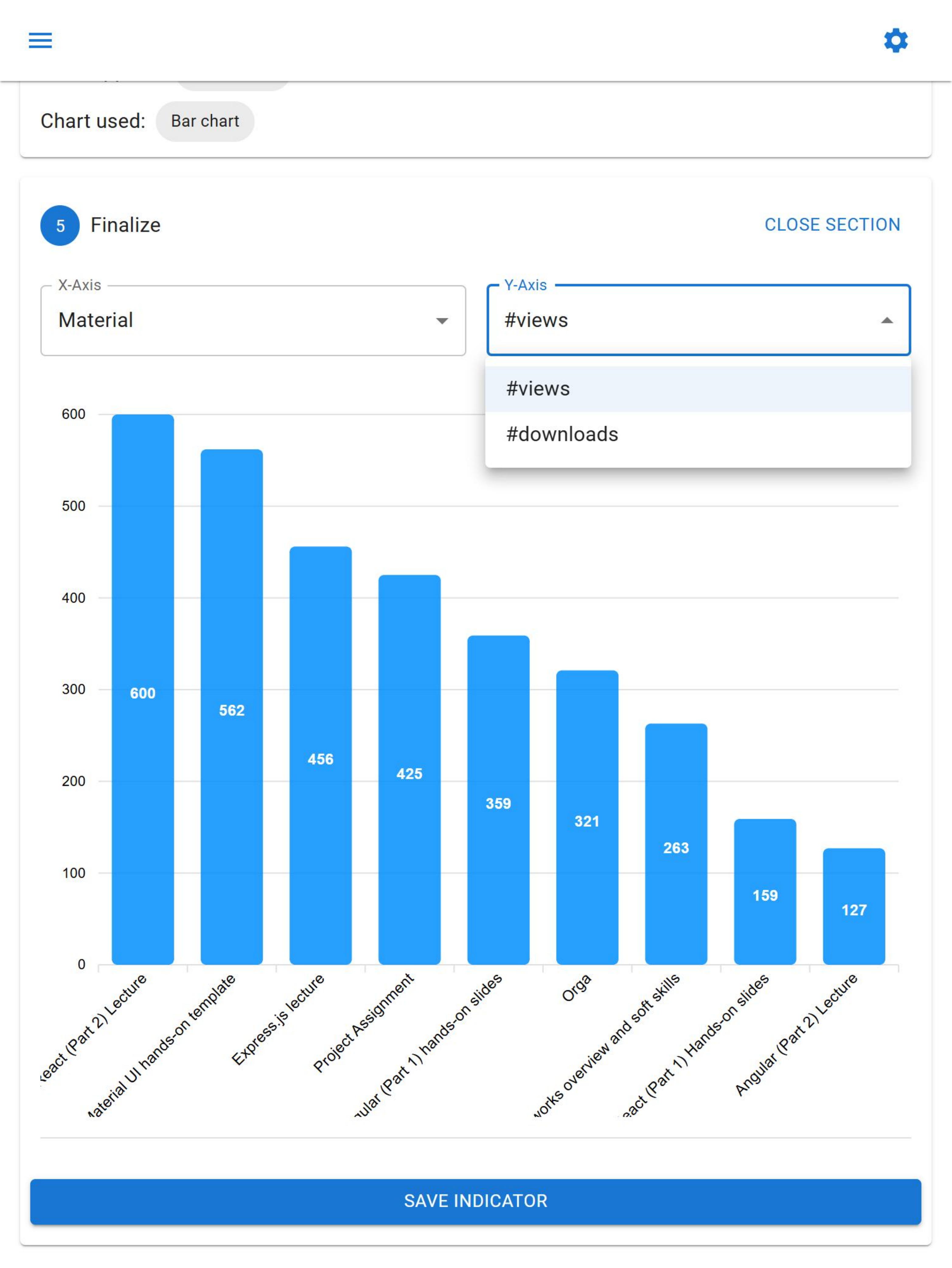}
            \caption{Finalize the indicator}
            \label{subfig:finalize-indicator}
        \end{subfigure}
        ~
        \begin{subfigure}[normla]{0.23\textwidth}
            \includegraphics[width=1\textwidth]{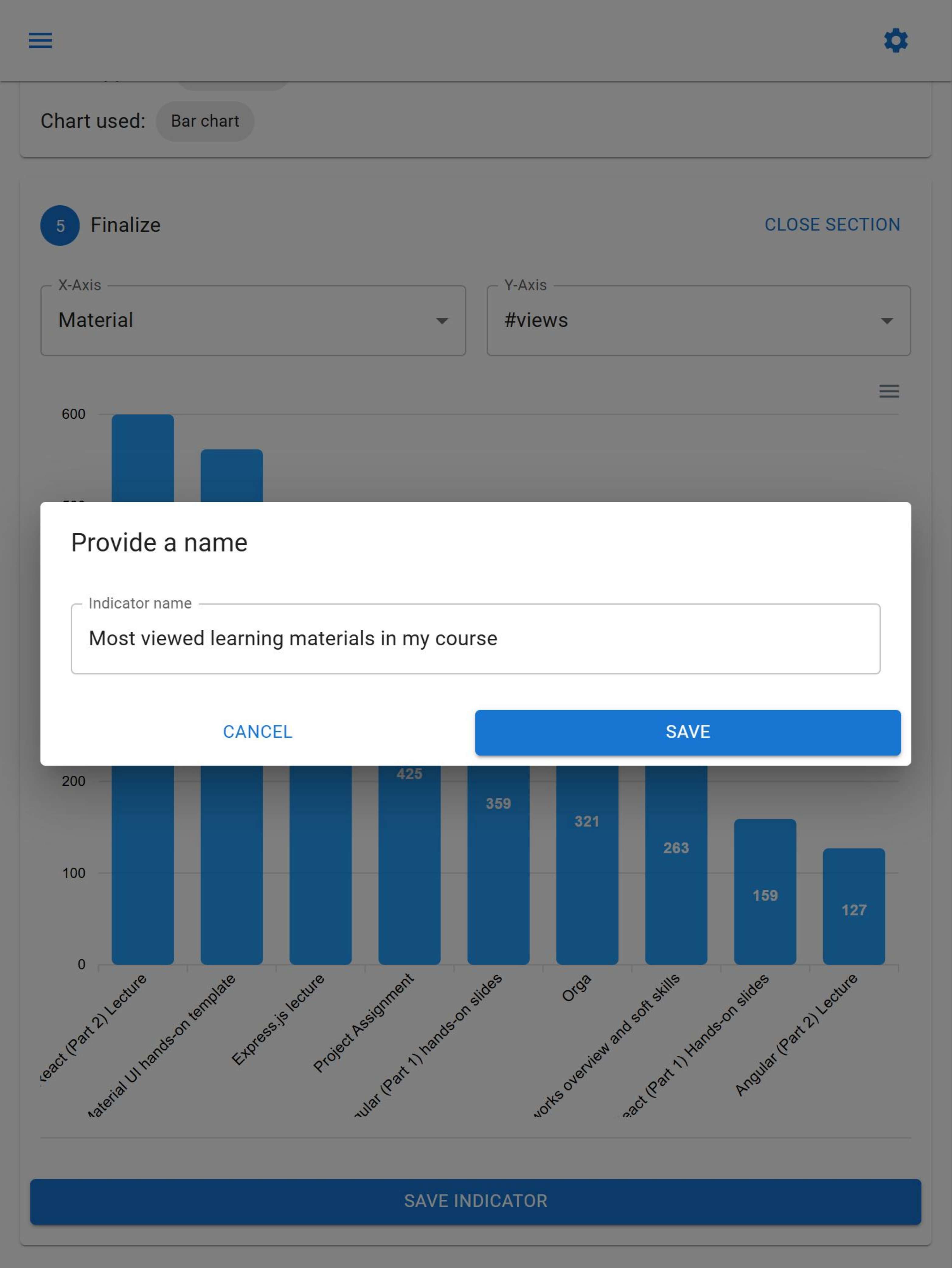}
            \caption{Name the indicator}
            \label{subfig:name-indicator}
        \end{subfigure}
    \end{center}
    \caption{Finalize indicator step}
    \label{fig:finalize-indicator-step}
\end{figure}

\begin{figure}[htpb]
    \begin{center}
        \begin{subfigure}[normla]{0.23\textwidth}
            \includegraphics[width=1\textwidth]{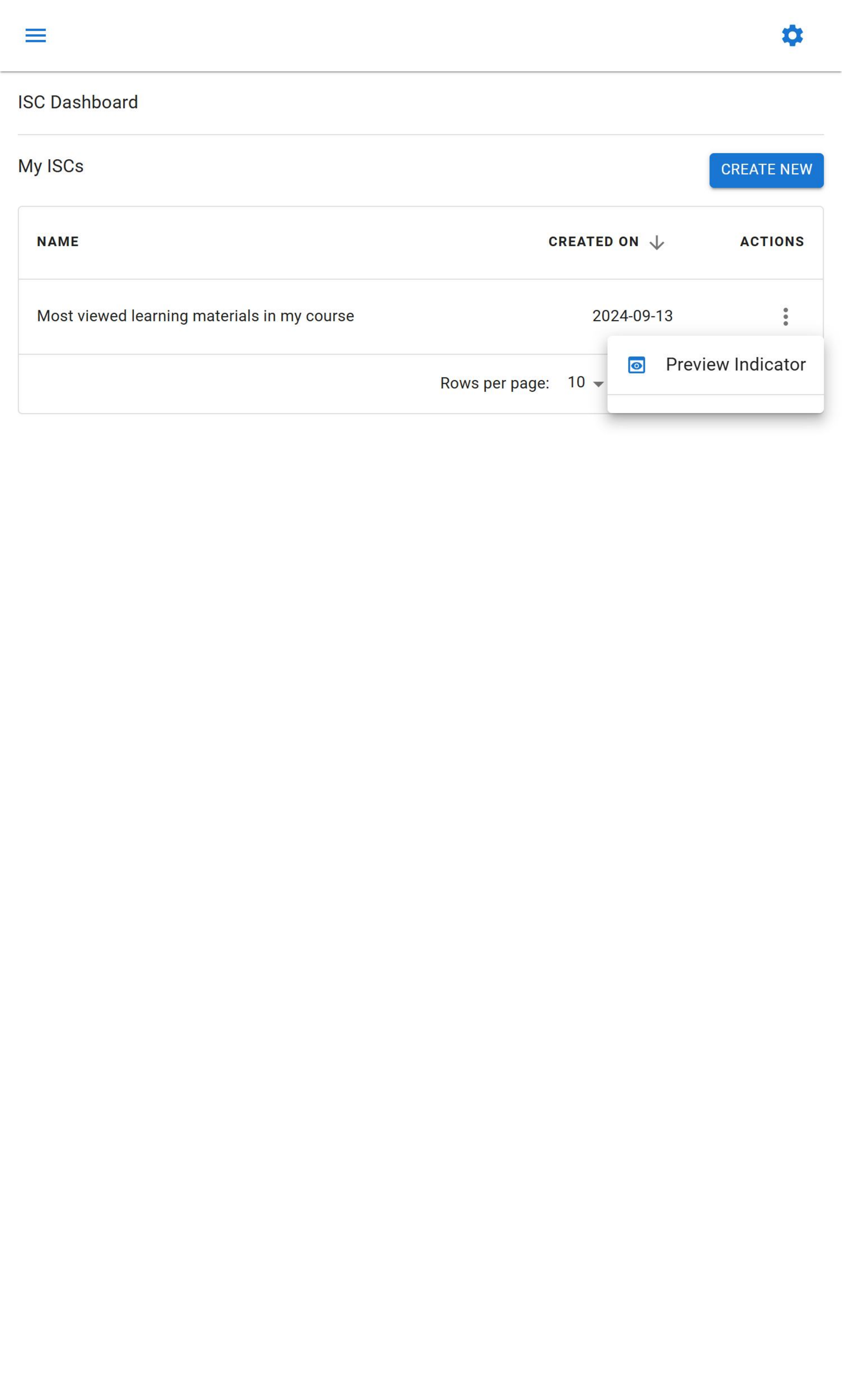}
            \caption{View ISC Dashboard}
            \label{subfig:dashboard}
        \end{subfigure}
        ~
        \begin{subfigure}[normla]{0.23\textwidth}
            \includegraphics[width=1\textwidth]{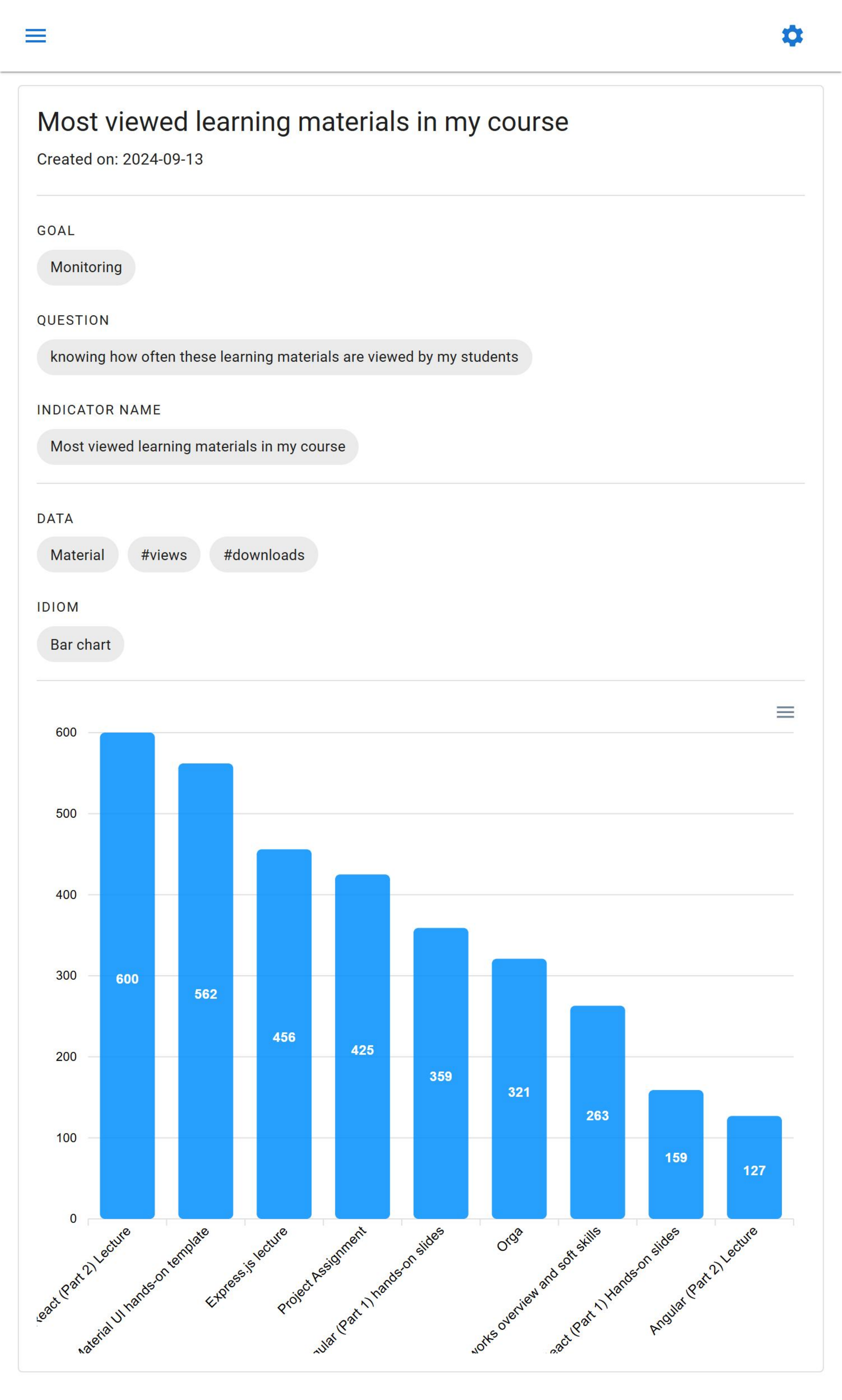}
            \caption{Preview ISC details}
            \label{subfig:preview}
        \end{subfigure}
    \end{center}
    \caption{Dashboard and preview page}
    \label{fig:dashboard-preview}
\end{figure}

%%%%%%%%%%%%%%%%%%%%%%%%%%%%%%%%%%%%%%%
Similar to the \textit{task-driven approach}, when a user clicks on \textit{Select Visualization}, they can directly choose an idiom/chart (Figure \ref{subfig:view-charts}) \textbf{(DG2)}. In this approach, the user will only get the recommendation of charts based on the data type specified in the `Specify your goal and question' UI section (Figure \ref{subfig:specify-goal-question}) \textbf{(DG3)}. Next, the user can click the `NEXT' button to select data (Figure \ref{subfig:create-data}), finalize the visualization (Figure \ref{fig:finalize-indicator-step}) by providing a name (Figure \ref{subfig:name-indicator}), and then save the ISC to the dashboard (Figure \ref{subfig:dashboard}). The user can also preview their ISC by opening the menu and clicking the `Preview Indicator' menu button to redirect to a preview page of the ISC (Figure \ref{subfig:preview}), which looks similar to the ISC example in Figure \ref{fig:isc-example}.
%%%%%%%%%%%%%%%%%%%%%%%%%%%%%%%%%%%%%%%
% Finalize

%%%%%%%%%%%%%%%%%%%%%%%%%%%%%%%%%%%%%%%
%%%%%%%%%%%%%%%%%%%%%%%%%%%%%%%%%%%%%%%
% Dashboard
%%%%%%%%%%%%%%%%%%%%%%%%%%%%%%%%%%%%%%%

%%%%%%%%%%%%%%%%%%%%%%%%%%%%%%%%%%%%%%%
\section{Evaluation}
\label{evaluation}
%%%%%%%%%%%%%%%%%%%%%%%%%%%%%%%%%%%%%%%
After systematically designing the \textit{ISC Creator}, we conducted a qualitative evaluation using a semi-structured interview format and a think-aloud protocol to gather in-depth feedback on the usage and attitudes towards an indicator designing process facilitated by the \textit{ISC Creator}. To evaluate the user acceptance of and satisfaction with the \textit{ISC Creator}, we used the technology acceptance model (TAM). The two most influential factors that describe users' intention to use a system are perceived usefulness (PU) and perceived ease of use (PEOU) \cite{davis1989perceived}. To note that by using the TAM questionnaire, we are not aiming at conducting a quantitative evaluation and generalizing our conclusions, but rather to use participants' answers to the TAM questionnaire as a starting point for a qualitative investigation of users' needs, preferences, and expectations from an LA design exercise, supported by interactive ISCs.  
% Since users’ perspectives on self-service LA are inherently subjective and individual, a quantitative approach would be insufficient for this goal, as this approach aims at analyzing empirical data for predetermined hypotheses. Thus, we deem a qualitative approach more appropriate to explore individual expectations from a self-service LA approach. 
%We do not aim to generalize our conclusions but to deeply understand the potential of human control to increase the transparency of LA and inform the future design of self-service LA environments. 

%%%%%%%%%%%%%%%%%%%%%%%%%%%%%%%%%%%%%%%
\subsection{Participants}
%%%%%%%%%%%%%%%%%%%%%%%%%%%%%%%%%%%%%%%
We conducted a user study with ten non-expert LA stakeholders (mainly students and teachers). We recruited participants via email, word-of-mouth, and social media to ensure a diverse sample across countries, educational levels, and study backgrounds. Six females and four males aged 18 and 44 completed the study. Two participants were invited from the low-fidelity prototype test phase. Most participants (n=8) reported being unfamiliar with LA and data analytics. Four participants reported being familiar with visualizations.  Most participants (n=8) were international students residing in Germany with sufficient English proficiency. The highest reported educational level was Masters (60\%), while nearly 40\% had a study background in Computer Science, and the rest were a mix of Business Intelligence, English, and Mathematics. All participants gave informed consent to participate in the study.

%%%%%%%%%%%%%%%%%%%%%%%%%%%%%%%%%%%%%%%
\subsection{Study Design}
%%%%%%%%%%%%%%%%%%%%%%%%%%%%%%%%%%%%%%%
Participants were first presented with an online survey via Google Forms, where they completed a questionnaire about their demographics and familiarity with LA, data analytics, and visualization. Next, they were introduced to the goals and concepts used in the \textit{ISC Creator}, with concrete examples provided for better understanding. We then conducted moderated think-aloud sessions where participants were asked to perform two tasks: (1) ``Assume you are a student, and you would like to create an indicator that shows the distribution of your grades and those of your classmates,'' and (2) ``Assume you are a teacher, and you would like to create a new indicator to show how active are the students in your course.'' 
Following the think-aloud method, participants were encouraged to verbalize any thoughts that came to mind during each interaction. Afterward, we conducted semi-structured interviews to gather in-depth feedback. These interviews were conducted online, lasted approximately 30 to 45 minutes, and were recorded with the participants' consent. During the interviews, participants were asked the following open-ended questions: (1) `What do you like the most about the current state of the \textit{ISC Creator}?', (2) `What do you like the least about the current state of the \textit{ISC Creator}?', (3) `Which parts or features of the \textit{ISC Creator} influenced your satisfaction with the tool? How?', and (4) `Do you have any suggestions for improving the \textit{ISC Creator}?'.
% (3) \textit{Do you have a sense of control when interacting with the \textit{ISC Creator}? How?}, (4) \textit{Does the interaction/controllability of the \textit{ISC Creator} influence the transparency of the application? How?}, (4.1) \textit{Which parts or features of the \textit{ISC Creator} give you a sense of transparency of the \textit{ISC Creator}? Why?}, (5) \textit{Does the interaction/controllability of the \textit{ISC Creator} influence your trust in the application? How?}, (5.1) \textit{Which parts or features of the \textit{ISC Creator} give you a sense of trust in the \textit{ISC Creator}? Why?}, 
After the semi-structured interviews, participants were also asked to fill out a questionnaire containing questions based on four constructs, namely \textbf{(1)} \textit{Perceived Usefulness} \cite{davis1989perceived}, \textbf{(2)} \textit{Perceived Ease of Use} \cite{davis1989perceived,pu2011user}, \textbf{(3)} \textit{Intention To Use} \cite{venkatesh2003user}, and \textbf{(4)} \textit{Satisfaction} \cite{pu2011user}, as shown in Table \ref{tab:constructs}. For each construct, answers were given on a 5-point Likert scale, ranging from 1 (``strongly disagree'') to 5 (``strongly agree'').
% \textbf{(4)} \textit{Control \& Personalization} \cite{pu2011user}, \textbf{(5)} \textit{Transparency} \cite{pu2011user,hellmann2022development}, \textbf{(6)} \textit{Trust} \cite{gefen2003trust,carter2005utilization}, 
\begin{table}
    \caption{Constructs and items}
    \label{tab:constructs}
    \centering
    %\begin{tabular}{| c | p{11cm} |}
    \begin{tabular}{| p{0.07\textwidth} | p{0.35\textwidth} |}
    \hline
    \textbf{Construct} & \textbf{Items} \\
    \hline
    \multirow{5}{*}{\shortstack[l]{Perceived \\ Usefulness}}
    & \textbf{pu1} - Using the \textit{ISC Creator} would enable me to generate the indicators more quickly. \\  \hhline{~-}
    & \textbf{pu2} - Using the \textit{ISC Creator} would enable me to generate the required indicators to support my learning/teaching activities. \\ \hhline{~-}
    & \textbf{pu3} - I would find the \textit{ISC Creator} a useful tool to create indicators. \\ \hhline{~-}
    & \textbf{pu4} - The \textit{ISC Creator} provided me with the interaction possibilities I expected from it.  \\ \hhline{~-}
    & \textbf{pu5} - Using the \textit{ISC Creator} would make it easier to create new indicators. \\ 
    \hline 
    \multirow{6}{*}{\shortstack[l]{Perceived \\ Ease of \\ Use}}
    & \textbf{peou1} - Learning to interact with the \textit{ISC Creator} would be easy for me. \\ \hhline{~-}
    & \textbf{peou2} - I would find it easy to get the \textit{ISC Creator} to do what I want it to do. \\ \hhline{~-}
    & \textbf{peou3} - I believe interacting with the \textit{ISC Creator} would be a clear and understandable process. \\ \hhline{~-}
    & \textbf{peou4} - I would find the \textit{ISC Creator} to be flexible to interact with.  \\ \hhline{~-}
    & \textbf{peou5} - I became familiar with the \textit{ISC Creator} very quickly. \\ \hhline{~-}
    & \textbf{peou6} - I would find the \textit{ISC Creator} easy to use. \\
    \hline
    \multirow{3}{*}{\shortstack[l]{Intention \\ To Use}}
    & \textbf{itu1} - I believe it is worthwhile to use the \textit{ISC Creator} to create indicators. \\ \hhline{~-}
    & \textbf{itu2} - Interacting with the \textit{ISC Creator} to create indicators is something I would do. \\ \hhline{~-}
    & \textbf{itu3} - I intend to use the \textit{ISC Creator} to create indicators in the future. \\ 
    % \hline
    % \multirow{3}{*}{\shortstack[l]{Control \& \\ Personalization}}
    % & \textbf{cp1} - I feel in control of indicator customization provided by the \textit{ISC Creator}. \\ \hhline{~-}
    % & \textbf{cp2} - The \textit{ISC Creator} allows me to modify the indicators based on my needs. \\ \hhline{~-}
    % & \textbf{cp3} - I am satisfied with the level of interactivity to support indicator customization provided by the \textit{ISC Creator}. \\
    % \hline
    % \multirow{3}{*} {Transparency}
    % & \textbf{tra1} - I understood how the indicators are created. \\ \hhline{~-}
    % & \textbf{tra2} - It was clear to me what kind of data the \textit{ISC Creator} uses to generate indicators. \\ \hhline{~-}
    % & \textbf{tra3} - I know what actions to perform in the \textit{ISC Creator} so that it generates indicators based on my needs. \\
    % \hline
    % \multirow{3}{*} {Trust}
    % & \textbf{tru1} - The \textit{ISC Creator} can be trusted to analyze learners’ data faithfully. \\ \hhline{~-}
    % & \textbf{tru2} - I can trust a learning analytics system that uses the indicators created with the \textit{ISC Creator}. \\ \hhline{~-}
    % & \textbf{tru3} - In my opinion, the \textit{ISC Creator} is trustworthy. \\
    \hline
    \multirow{1}{*} {Satisfaction}
    & \textbf{sa1} - Overall, I am satisfied with the \textit{ISC Creator}. \\
    \hline
    \end{tabular}
\end{table}
% \begin{figure}[!ht]
% 	\centering
% 	\includegraphics[width=0.47\textwidth]{chapters/4-evaluation/images/evaluation-constructs.pdf}\\
% 	\caption{Constructs and items}
% 	\label{tab:constructs}
% \end{figure}
%%%%%%%%%%%%%%%%%%%%%%%%%%%%%%%%%%%%%%%

%%%%%%%%%%%%%%%%%%%%%%%%%%%%%%%%%%%%%%%
\subsection{Analysis and Results}
The feedback was divided into four main dimensions: \textit{Perceived Usefulness}, \textit{Perceived Ease of Use}, \textit{Intention to Use}, and \textit{Satisfaction}.

%%%%%%%%%%%%%%%%%%%%%%%%%%%%%%%%%%%%%%%
\begin{figure}[!ht]
	\centering
	\includegraphics[width=0.49\textwidth]{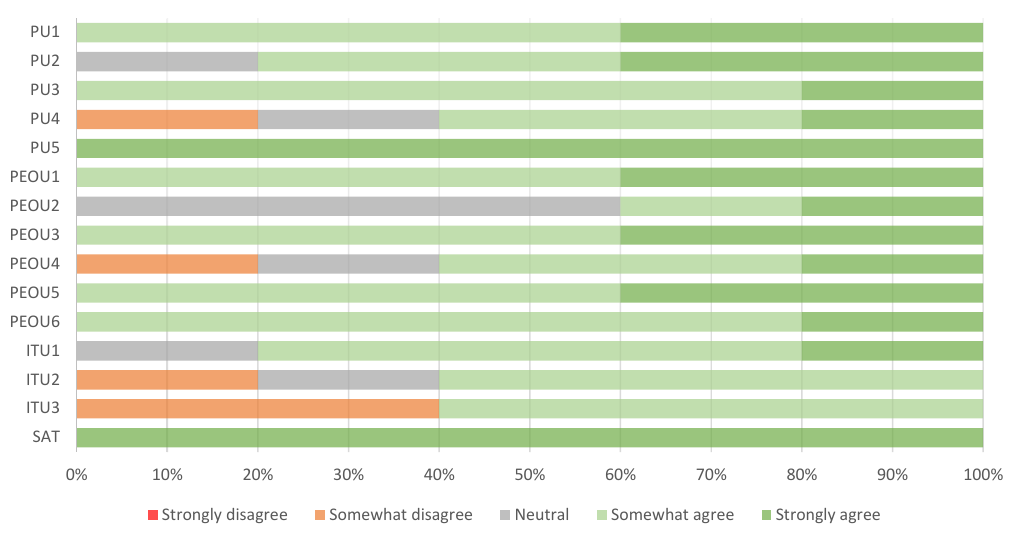}\\
	\caption{Results from the questionnaires: Acceptance \& User Satisfaction }
	\label{fig:results-questionnaire}
\end{figure}
%%%%%%%%%%%%%%%%%%%%%%%%%%%%%%%%%%%%%%%

%%%%%%%%%%%%%%%%%%%%%%%%%%%%%%%%%%%%%%%
\subsubsection{Perceived Usefulness}
%%%%%%%%%%%%%%%%%%%%%%%%%%%%%%%%%%%%%%%
It refers to how individuals believe using a particular system would enhance their work performance \cite{davis1989perceived}. To measure the perceived usefulness of the \textit{ISC Creator}, Table \ref{tab:constructs} (pu1-pu5) presents the questionnaire items. As indicated in Figure \ref{fig:results-questionnaire}, most of the participants agreed that they could quickly design LA indicators using the \textit{ISC Creator} and that it could facilitate their teaching/learning activities. Furthermore, after using the \textit{ISC Creator}, they perceived it as a beneficial tool for designing LA indicators. For instance, \textit{\textbf{P2:} ``The system's workflow is easy for me as a teacher because I am unfamiliar with the concept of indicators, and the system guides me step-by-step''}, \textit{\textbf{P3:} ``The system provides me with chart descriptions and a preview before my final decision, and I can see the requirements for using any chart''}, and \textit{\textbf{P5:} ``I can generate an indicator with just a few clicks.''} Participants also had positive opinions regarding the interaction possibilities provided by The \textit{ISC Creator}, as mentioned by \textit{\textbf{P3:} ``I could choose a chart, upload the data, and manipulate my data. The interaction with the system is extremely straightforward, and I could quickly create an indicator''.}. However, two participants mentioned that the limitation of available visualization hindered them from designing an ISC, \textit{\textbf{P7:} ``There are not enough types of visualizations available to realize my imagined indicator, such as a gauge type indicator to reflect the performance.''} A few participants (n=3) additionally mentioned that they expected more customization options, such as hiding and changing the position of the legends.
%However, some participants felt that they lacked control over the system because it did not display the values of one of their created data columns, as mentioned by \textit{``\textbf{P2:} I was unable to see the column value in the indicator, and I did not understand why. It was beyond my control to fix it''}.
% Some participants expressed a positive attitude towards the effect of interaction and control on perceived transparency: \textit{``\textbf{P3:} The system was able to show me everything I needed, and I was not lost in finding the appropriate interaction elements such as buttons or searching to learn what to do next.''}, \textit{``\textbf{P5:} When selecting the x-axis and y-axis of my chart, I could see the column names appropriate for the axis. Ultimately, I get the results based on my data''}.

%%%%%%%%%%%%%%%%%%%%%%%%%%%%%%%%%%%%%%%
\subsubsection{Perceived Ease of Use}
%%%%%%%%%%%%%%%%%%%%%%%%%%%%%%%%%%%%%%%
It refers to the extent to which an individual believes that using a particular system requires minimal effort \cite{davis1989perceived}. The questionnaire items to measure the perceived ease of use of the \textit{ISC Creator} are listed in Table \ref{tab:constructs} (peou1-peou6). As illustrated in Figure \ref{fig:results-questionnaire}, the participants acknowledged that the \textit{ISC Creator} was generally intuitive, \textit{\textbf{P1:} ``The dataset column and types provided an understanding of the requirements for the charts''}, \textit{\textbf{P5:} ``I did not expect, but I found it helpful that the dataset gets automatically populated with the name of the column I specified earlier.''} However, they also identified some areas for improvement, such as avoiding repetitive steps and providing more customization possibilities. For instance, \textit{\textbf{P1:} ``I don't like creating one row at a time. I prefer to create multiple rows with one click. Furthermore, I cannot change the color of my charts.''} 
% TODO: 
% Additionally, one participant (\textbf{P2}) suggested that a preview panel in the dashboard could avoid going into the edit mode.
%, as mentioned by \textit{``\textbf{P2:} In the dashboard, I had to click the edit button to view my indicator. I want to see a preview panel to avoid going into the edit mode.''}
%%%%%%%%%%%%%%%%%%%%%%%%%%%%%%%%%%%%%%%
\subsubsection{Intention to Use}
%%%%%%%%%%%%%%%%%%%%%%%%%%%%%%%%%%%%%%%
The intention to use (ITU) technology refers to the extent to which a person intends to adopt and use it \cite{venkatesh2003user}. Table \ref{tab:constructs} (itu1-itu3) presents the questionnaire items used to measure users' intention to use the ISC. As shown in Figure \ref{fig:results-questionnaire}, some participants (n=4) expressed reluctance to use the system in the future, primarily due to the limited customization options available for charts. Additionally, some users were confused by the lack of system feedback reflecting changes in the chart. For instance, one participant noted, \textit{\textbf{P7:} ``I can imagine that if I have an important, large dataset for a conference, and if I make a small change in the data, I don't get feedback on whether anything has changed in the chart.''} A few participants (n=2) found the absence of an export functionality problematic, questioning the \textit{ISC Creator}'s utility, stating, \textit{\textbf{P9:} ``I don't see why I would use this if I can't export my indicators and share them elsewhere.''} Moreover, one participant mentioned that the \textit{ISC Creator} preview page should also show the data in tabular form. Furthermore, some participants (n=4) highlighted the absence of privacy settings as a concern affecting their trust in the system, stating, \textit{\textbf{P4:} ``I want the system to confirm that my uploaded data will not be used elsewhere.''}

% The intention to use (ITU) technology refers to the extent to which a person intends to use it \cite{venkatesh2003user}. Table \ref{tab:constructs} (itu1-itu3) presents the questionnaire items used to measure users' intention to use the ISC. Figure \ref{fig:results-questionnaire}c indicates that some participants disagreed with the system in the future due to the limited customization available for the charts. Some users expressed confusion due to lack of feedback from the system that reflected in the chart, which led to a decrease in trust, as noted by \textit{``\textbf{P7:} I can imagine that if I have an important big dataset for a conference, and if I make a small change in the data, I do not get feedback whether something has changed in the chart.''} Additionally, one participant mentioned that the absence of privacy settings would affect their trust in the system, stating \textit{``\textbf{P4:} I want the system to mention that my uploaded data will not be used anywhere else.''}

%%%%%%%%%%%%%%%%%%%%%%%%%%%%%%%%%%%%%%%
\subsubsection{Satisfaction}
%%%%%%%%%%%%%%%%%%%%%%%%%%%%%%%%%%%%%%%
Assessing users' overall satisfaction is crucial to determining their thoughts and feelings using the \textit{ISC Creator} \cite{pu2011user}. Table \ref{tab:constructs} (sa1) presents the questionnaire item used to measure the overall satisfaction of the \textit{ISC Creator}. All participants expressed high satisfaction with the \textit{ISC Creator} (Figure \ref{fig:results-questionnaire}). When we concretely asked about which features influenced their satisfaction with the tool, participants expressed their satisfaction in various ways. Some were satisfied with the interaction and control mechanisms provided by the \textit{ISC Creator}. For instance, \textit{\textbf{P3:} ``I like that I can customize the charts to fit my needs''} and \textit{\textbf{P2:} ``I can control the application by creating my dataset and selecting appropriate data column types, and then I was able to use the data to visualize in the chart. Everything was under my control.''} Other participants were satisfied with the feedback and recommendations provided by the \textit{ISC Creator}, as evidenced by comments such as \textit{\textbf{P6:} ``I was able to see my indicator that matched my data''}, \textit{\textbf{P10:} ``The system recommends charts based on my data types, and it is correct when I think about it by myself''}, \textit{``\textbf{P4:} The system was able to show me everything I needed, and I was not lost in finding the appropriate interaction elements such as buttons or searching to learn what to do next''}, and \textit{\textbf{P5:} ``When selecting the x-axis and y-axis of my chart, I could see the column names appropriate for the axis. Ultimately, I get the results based on my data.''}
%%%%%%%%%%%%%%%%%%%%%%%%%%%%%%%%%%%%%%%
\section{Limitations}
\label{limitations}
%%%%%%%%%%%%%%%%%%%%%%%%%%%%%%%%%%%%%%%
Some limitations of this work need to be addressed. First, although we engaged a diverse group of non-expert LA stakeholders from the local university, including a broader and more varied population would better assess the \textit{ISC Creator}'s effectiveness among end-users with diverse backgrounds and experiences. Additionally, we conducted a qualitative user study with only n=10 participants. As a result, the findings should be interpreted cautiously and cannot be generalized. A quantitative study with a larger sample size would likely yield more significant and reliable results.
%%%%%%%%%%%%%%%%%%%%%%%%%%%%%%%%%%%%%%%
\section{Conclusion and Future Work}
\label{conclusion}
%%%%%%%%%%%%%%%%%%%%%%%%%%%%%%%%%%%%%%%
In this paper, we aimed to provide an effective and efficient method for creating low-fidelity learning analytics (LA) indicators using Indicator Specification Cards (ISC). To this end, we presented the systematic design, implementation, and evaluation of the \textit{ISC Creator}, which enables the low-cost and flexible design of LA indicators through a human-centered design approach and adherence to information visualization guidelines. Additionally, we introduced three flexible approaches, namely task-driven approach, data-driven approach, and visualization-driven approach, that allow users to customize their LA indicators according to their needs and goals. Based on the technology acceptance model, we conducted a qualitative evaluation of user acceptance and satisfaction with the \textit{ISC Creator}. The results indicated that empowering users with control over their indicator design process and providing theoretically grounded recommendations can enhance the acceptance and adoption of LA tools. While we acknowledge that our findings may not be generalizable, they offer valuable insights for designing interactive, human-centered LA tools. Future research will include quantitative studies to investigate further the effects of the \textit{ISC Creator} on users' perceptions of control, transparency, and trust, as well as the interaction effects between these important user-centered aspects.
% This paper aimed to provide an effective and efficient method for creating low-fidelity learning analytics (LA) indicators using Indicator Specification Cards (ISC). We presented the design, implementation, and evaluation details of the ISC Creator that allows low-cost and flexible design of LA indicators, following a Human-Centered Design (HCD) approach and applying information visualization guidelines. Moreover, we proposed three flexible approaches, namely \textit{task-driven approach}, \textit{data-driven approach}, and \textit{visualization-driven approach}, to enable users to customize their LA indicators according to their needs. Furthermore, based on the technology acceptance model, we conducted a qualitative evaluation of the user acceptance and satisfaction with the ISC Creator. The evaluation results showed that giving control to users in their interaction indicator design process and supporting them with theoretically sound recommendations can push forward the acceptance and adoption of LA tools. Although we acknowledge that our results may not be generalizable, we believe that they serve as valuable reference points for designing interactive human-centered LA tools. In future research, we plan to conduct quantitative studies to examine further the effects of the ISC Creator on users' perceptions of transparency and trust.

\bibliographystyle{plainnat}
\bibliography{sample-bibliography} 

\begin{thebibliography}{20}
\providecommand{\natexlab}[1]{#1}
\providecommand{\url}[1]{\texttt{#1}}
\expandafter\ifx\csname urlstyle\endcsname\relax
  \providecommand{\doi}[1]{doi: #1}\else
  \providecommand{\doi}{doi: \begingroup \urlstyle{rm}\Url}\fi

\bibitem[Alvarez et~al.(2020)Alvarez, Martinez-Maldonado, and Shum]{alvarez2020deck}
Carlos~Prieto Alvarez, Roberto Martinez-Maldonado, and Simon~Buckingham Shum.
\newblock La-deck: A card-based learning analytics co-design tool.
\newblock In \emph{Proceedings of the tenth international conference on learning analytics \& knowledge}, pages 63--72, 2020.

\bibitem[Braun and Clarke(2006)]{braun2006using}
Virginia Braun and Victoria Clarke.
\newblock Using thematic analysis in psychology.
\newblock \emph{Qualitative research in psychology}, 3\penalty0 (2):\penalty0 77--101, 2006.

\bibitem[Buckingham~Shum et~al.(2019)Buckingham~Shum, Ferguson, and Martinez-Maldonado]{buckingham2019human}
Simon Buckingham~Shum, Rebecca Ferguson, and Roberto Martinez-Maldonado.
\newblock Human-centred learning analytics.
\newblock \emph{Journal of Learning Analytics}, 6\penalty0 (2):\penalty0 1--9, 2019.

\bibitem[Chatti and Muslim(2019)]{chatti2019perla}
Mohamed~Amine Chatti and Arham Muslim.
\newblock The perla framework: Blending personalization and learning analytics.
\newblock \emph{International review of research in open and distributed learning}, 20\penalty0 (1), 2019.

\bibitem[Chatti et~al.(2020)Chatti, Muslim, Guesmi, Richtscheid, Nasimi, Shahin, and Damera]{chatti2020design}
Mohamed~Amine Chatti, Arham Muslim, Mouadh Guesmi, Florian Richtscheid, Dawood Nasimi, Amin Shahin, and Ritesh Damera.
\newblock How to design effective learning analytics indicators? a human-centered design approach.
\newblock In \emph{European conference on technology enhanced learning}, pages 303--317. Springer, 2020.

\bibitem[Chatti et~al.(2021)Chatti, Y{\"u}cepur, Muslim, Guesmi, and Joarder]{chatti2021designing}
Mohamed~Amine Chatti, Volkan Y{\"u}cepur, Arham Muslim, Mouadh Guesmi, and Shoeb Joarder.
\newblock Designing theory-driven analytics-enhanced self-regulated learning applications.
\newblock In \emph{Visualizations and Dashboards for Learning Analytics}, pages 47--68. Springer, 2021.

\bibitem[Davis(1989)]{davis1989perceived}
Fred~D Davis.
\newblock Perceived usefulness, perceived ease of use, and user acceptance of information technology.
\newblock \emph{MIS quarterly}, pages 319--340, 1989.

\bibitem[Dimitriadis et~al.(2021)Dimitriadis, Mart{\'\i}nez-Maldonado, and Wiley]{dimitriadis2021human}
Yannis Dimitriadis, Roberto Mart{\'\i}nez-Maldonado, and Korah Wiley.
\newblock Human-centered design principles for actionable learning analytics.
\newblock In \emph{Research on E-Learning and ICT in Education}, pages 277--296. Springer, 2021.

\bibitem[Dollinger et~al.(2019)Dollinger, Liu, Arthars, and Lodge]{dollinger2019working}
Mollie Dollinger, Danny Liu, Natasha Arthars, and Jason~M Lodge.
\newblock Working together in learning analytics towards the co-creation of value.
\newblock \emph{Journal of Learning Analytics}, 6\penalty0 (2):\penalty0 10--26, 2019.

\bibitem[Ga{\v{s}}evi{\'c} et~al.(2015)Ga{\v{s}}evi{\'c}, Dawson, and Siemens]{gavsevic2015let}
Dragan Ga{\v{s}}evi{\'c}, Shane Dawson, and George Siemens.
\newblock Let’s not forget: Learning analytics are about learning.
\newblock \emph{TechTrends}, 59\penalty0 (1):\penalty0 64--71, 2015.

\bibitem[Ga{\v{s}}evi{\'c} et~al.(2016)Ga{\v{s}}evi{\'c}, Dawson, and Pardo]{gavsevic2016we}
Dragan Ga{\v{s}}evi{\'c}, Shane Dawson, and Abelardo Pardo.
\newblock How do we start? state directions of learning analytics adoption.
\newblock 2016.

\bibitem[Holstein et~al.(2019)Holstein, McLaren, and Aleven]{holstein2019co}
Kenneth Holstein, Bruce~M McLaren, and Vincent Aleven.
\newblock Co-designing a real-time classroom orchestration tool to support teacher--ai complementarity.
\newblock \emph{Journal of Learning Analytics}, 6\penalty0 (2), 2019.

\bibitem[Jivet et~al.(2018)Jivet, Scheffel, Specht, and Drachsler]{jivet2018license}
Ioana Jivet, Maren Scheffel, Marcus Specht, and Hendrik Drachsler.
\newblock License to evaluate: Preparing learning analytics dashboards for educational practice.
\newblock In \emph{Proceedings of the 8th international conference on learning analytics and knowledge}, pages 31--40, 2018.

\bibitem[Klerkx et~al.(2017)Klerkx, Verbert, and Duval]{klerkx2017learning}
Joris Klerkx, Katrien Verbert, and Erik Duval.
\newblock Learning analytics dashboards.
\newblock 2017.

\bibitem[Munzner(2014)]{munzner2014visualization}
Tamara Munzner.
\newblock \emph{Visualization analysis and design}.
\newblock CRC press, 2014.

\bibitem[Muslim et~al.(2017)Muslim, Chatti, Mughal, and Schroeder]{muslim2017goal}
Arham Muslim, Mohamed~Amine Chatti, Memoona Mughal, and Ulrik Schroeder.
\newblock The goal-question-indicator approach for personalized learning analytics.
\newblock In \emph{CSEDU (1)}, pages 371--378, 2017.

\bibitem[Norman(2013)]{norman2013design}
Don Norman.
\newblock \emph{The design of everyday things: Revised and expanded edition}.
\newblock Basic books, 2013.

\bibitem[Pu et~al.(2011)Pu, Chen, and Hu]{pu2011user}
Pearl Pu, Li~Chen, and Rong Hu.
\newblock A user-centric evaluation framework for recommender systems.
\newblock In \emph{Proceedings of the fifth ACM conference on Recommender systems}, pages 157--164, 2011.

\bibitem[Sarmiento and Wise(2022)]{sarmiento2022participatory}
Juan~Pablo Sarmiento and Alyssa~Friend Wise.
\newblock Participatory and co-design of learning analytics: An initial review of the literature.
\newblock In \emph{LAK22: 12th International Learning Analytics and Knowledge Conference}, pages 535--541, 2022.

\bibitem[Venkatesh et~al.(2003)Venkatesh, Morris, Davis, and Davis]{venkatesh2003user}
Viswanath Venkatesh, Michael~G Morris, Gordon~B Davis, and Fred~D Davis.
\newblock User acceptance of information technology: Toward a unified view.
\newblock \emph{MIS quarterly}, pages 425--478, 2003.

\end{thebibliography}
\end{document}